\documentclass[traditabstract]{aa}
\usepackage{graphicx}
\usepackage{txfonts}
\usepackage{natbib}
\bibpunct{(}{)}{;}{a}{}{,}
\begin{document}
    \title{
    The Polish doughnuts revisited
    }
   \subtitle{
   I. The angular momentum distribution and equipressure surfaces
            }
   \author{
              Qian Lei \inst{1,2}
          \and
              Marek A. Abramowicz \inst{1, 3}
          \and
             P. Chris Fragile \inst{4,1}
          \and
             Ji{\v r}{\'{\i}} Hor{\'a}k \inst{5,1}
          \and
             Mami Machida \inst{6}
          \and
             Odele Straub \inst{3,1}
          }
   \institute{
             Department of Physics, G\"oteborg University,
             SE-412-96 G\"oteborg, Sweden    \\
             \email{Marek.Abramowicz@physics.gu.se}
         \and
             Department of Astronomy, Peking University Cheng Fu St. 209,
             100871 Beijing,
             China\thanks{permanent address} \\
             \email{qianlivan@gmail.com}
         \and
             N. Copernicus Astronomical Center, Polish Academy
             of Sciences,
             Bartycka 18, 00-716 Warszawa,
             Poland \\
             \email{odele@camk.edu.pl}
         \and
             Physics \& Astronomy College of Charleston,
             58 Coming Street Charleston SC
             29424, U.S.A. \\
             \email{FragileP@cofc.edu}
         \and
            Astronomical Institute, Academy of Sciences of the
            Czech Republic,
            Bo{\v c}ni II/1401a, 141-31  Prague,
            Czech Republic
            \\
            \email{jiri.horak@cdsw.cz}
         \and
            National Astronomical Observatory of Japan 2-21-1 Osawa,
            Mitaka, 181-8588 Tokyo, Japan\\
            \email{mami@th.nao.ac.jp}
             }
\date{Received ????; accepted ???? }
  \abstract{We construct a new family of analytic models of black
   hole accretion disks in dynamical equilibria.
   Our construction
   is based on assuming distributions of angular momentum and entropy.
   For a particular choice of the distribution of angular momentum,
   we calculate the shapes of equipressure surfaces.
   The equipressure surfaces we find
   are similar to those in thick, slim and thin disks, and to
   those in ADAFs.}
\authorrunning{Qian Lei, M.A.\,Abramowicz, P.C.\,Fragile,
               J.\,Hor{\'a}k, M.\,Machida \& O.\,Straub}
\titlerunning{The Polish doughnuts revisited}
  \maketitle

\section{Introduction}


In accretion disk theory one is often interested in phenomena that
occur on a ``dynamical'' timescale ${\cal T}_{0}$ much shorter
than the ``viscous'' timescale ${\cal T}[{\cal L}]$ needed for
angular momentum redistribution and the ``thermal'' timescale ${\cal
T}[{\cal S}]$ needed for entropy redistribution\footnote{We use the spherical Boyer-Lindquist coordinates $t, \phi, r, \theta$, the geometrical units $c$ $=$ $1$ $=$ $G$ and the $+---$ signature. The Kerr metric is described by the ``geometrical'' mass $M$ and the ``geometrical'' spin parameter $0 < a < 1$, that relate to the ``physical'' mass and angular momentum by the rescaling, $M = GM_{\rm phys}/c^2$, $a = J_{\rm phys}/(M\,c)$. Partial derivatives are denoted by $\partial_i$ and covariant derivatives by $\nabla_i$.},
\begin{equation}
\label{timescales} {\cal T}_{0} \ll {\rm min}\left({\cal
T}[{\cal L}], {\cal T}[{\cal S}] \right).
\end{equation}
The question whether it is physically legitimate to approximately
describe the black hole accretion flows (at least in some
``averaged'' sense) in terms of stationary (independent on $t$)
and axially symmetric (independent on $\phi$) dynamical
equilibria, is not yet resolved. While observations seem to suggest that many real
astrophysical sources experience periods in which this assumption
is quite reasonable, several authors point out that the results of
recent numerical simulations seem to indicate that the MRI and
other instabilities make the black hole accretion flows genuinely
non-steady and non-symmetric, and that the very concept of the
separate timescales (\ref{timescales}) may be questionable in the
sense that locally ${\cal T}_{0} \approx {\cal T}[{\cal L}]
\approx {\cal T}[{\cal S}]$. However, this assumption has been
made in {\it all} existing comparisons between theory and
observations, be they by detailed spectral fitting
\citep[e.g.][and references there]{sha-2007, sha-2008}, line
profile fitting \citep[e.g.][]{fab-2003}, or studying small
amplitude oscillations \citep[see][for references]{abr-2005}. It
seems that the present understanding of the black hole accretion
phenomenon rests, in a major way, on studies of stationary and
axially symmetric models.

From the point of view of mathematical self-consistency, in
modeling of these stationary and axially symmetric dynamical
equilibria, distributions of the {\it conserved} angular momentum
and entropy,
\begin{equation}
\label{distribution-lagrangian} \ell = \ell(\xi, \eta),~~~ s =
s(\xi, \eta),
\end{equation}
may be considered as being {\it free functions} of the Lagrangian
coordinates \citep{ost-1970, abr-1970, bar-1970}. The Lagrangian
coordinates $\xi, \eta$ are defined by demanding that a narrow
ring of matter $(\xi, \xi + d\xi)$, $(\eta, \eta + d\eta)$ has the
rest mass $dM_0 = \rho_0(\xi, \eta)d\xi d\eta$ with $\rho_0$ being
the rest mass density. In the full physical description, the form
of the functions in (\ref{distribution-lagrangian}) is not
arbitrary but given by the dissipative processes, like viscosity
and radiative transfer. At present, several important aspects of
these processes are still unknown, so there is still no practical
way to calculate physically consistent models of accretion flows
from first principles, without involving some ad hoc assumptions,
or neglecting some important processes. Neither the hydrodynamical
simulations (that e.g. use the ad hoc $\alpha\,=\,$const viscosity
prescription), nor the present day MHD simulations (that e.g.
neglect radiative transfer) could be considered satisfactory.
Furthermore, the simplifications made in these simulations are
mathematically equivalent to guessing free functions (such as the
entropy distribution).
Bohdan Paczy{\'n}ski pointed out that it could often be
more pragmatic to make a physically motivated guess of the final
result, e.g. to guess the form of the angular momentum and entropy
distributions.

In practice, it is far easier to guess and use the coordinate
distributions of the {\it specific} angular momentum and entropy,
\begin{eqnarray}
 {\cal L} &=& {\cal L}(r, \theta),
 \label{momentum-distribution}\\
 {\cal S} &=& {\cal S}(r, \theta),
 \label{entropy-distribution}
\end{eqnarray}
than the Lagrangian distributions (\ref{distribution-lagrangian}).
However, one does not known a priori the relation between the
conserved $\ell$ and specific ${\cal L}$ angular momenta (and
entropy), or the functions,  $\xi = \xi(r,\theta)$, $\eta =
\eta(r, \theta)$. Thus, assuming (\ref{momentum-distribution}) and
(\ref{entropy-distribution}) is not equivalent to assuming
(\ref{distribution-lagrangian}), and usually it should be a
subject to some consistency conditions. We shall return to this
point in Section \ref{discussion}.

In several ``astrophysical scenarios'' one indeed guesses a
particular form of (\ref{momentum-distribution}) and
(\ref{entropy-distribution}). For example, the celebrated
\citet{sha-sun-1974} {\it thin disk} model assumes the Keplerian
distribution of angular momentum,
\begin{equation}
\label{Keplerian} {\cal L}(r, \theta) = {\cal L}_K(r) \equiv
\frac{M^{1/2}\,\left(r^2 - 2aM^{1/2}r^{1/2} + a^2\right)} {r^{3/2}
- 2Mr^{1/2} + aM^{1/2}},
\end{equation}
and the popular {\it cold-disk-plus-hot-corona} model assumes a
low entropy flat disk surrounded by high entropy, more spherical
corona. These models contributed considerably to the understanding
of black-hole accretion physics.

The mathematically simplest assumption for the angular momentum
and entropy distribution is, obviously,
\begin{eqnarray}
 {\cal L}(r, \theta) &=& {\cal L}_0 = {\rm const},
 \label{constant-momentum}
\\
{\cal S}(r, \theta) &=& {\cal S}_0 = {\rm const}.
\label{constant-entropy}
\end{eqnarray}
This was used by Paczy{\'n}ski and his Warsaw team to introduce the
{\it thick disk} models \citep{abr-1978, koz-1978, jar-1980,
pac-wii-1980, abr-1980, abr-1981,
pac-1982}. Thick disks have characteristic
toroidal shapes, resembling a doughnut. Probably for this reason,
Martin Rees coined the name of {\it Polish
doughnuts}\footnote{However, real Polish doughnuts
(called {\it p{\c a}czki} in Polish) have spherical shapes. They
are definitely non-toroidal
--- see e.g. http://en.wikipedia.org/wiki/Paczki ~.} for them.

\begin{figure}
  \centering
   \includegraphics[width=9cm]{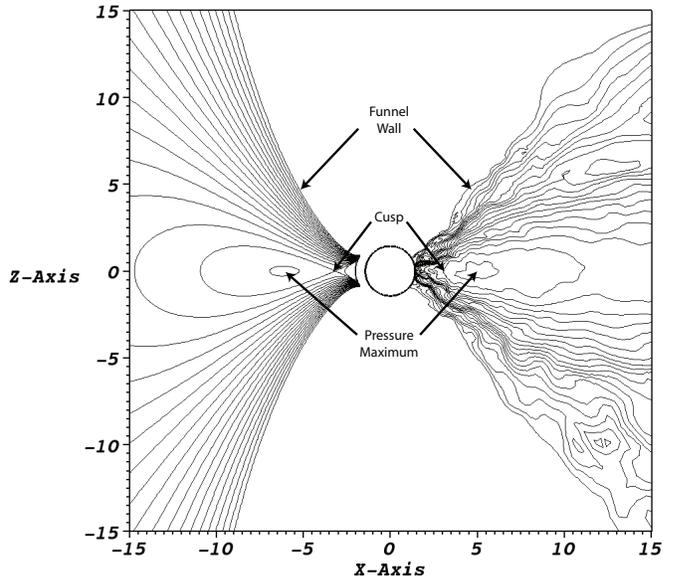}
      \caption{Equipressure surfaces in a very simple and analytic
      Polish doughnut (left, with linear spacing), and a sophisticated,
      state-of-art
      full 3D MHD numerical simulation (right, with logarithmic
      spacing). Although the shapes of equipressure surfaces are
      remarkably similar, in the numerical model the pressure
      gradient is seriously larger, and visibly enhanced along roughly
      conical surfaces, approximately $30^{\circ}$ from the
      equatorial plane. \citep[Figure taken from][]{abr-fra-2008}}
         \label{analytic-numerical}
   \end{figure}

Figure \ref{analytic-numerical} shows a comparison of a
state-of-art MHD simulation of black-hole accretion (time and
azimuth averaged) with a Polish doughnut corresponding to a
particular ${\cal L}_0$. Both models show the same characteristic
features of black hole accretion: (i) a funnel along the rotation
axis, relevant for jet collimation and acceleration; (ii) a
pressure maximum, possibly relevant for epicyclic oscillatory
modes; and (iii) a cusp-like self-crossing of one particular
equipressure surface, relevant for an inner boundary condition,
and for stabilization of the Papaloizou-Pringle \citep{bla-1987},
thermal, and viscous instabilities \citep{abr-1971}. The cusp is
located between the radii of marginally stable and marginally
bound circular orbits,
\begin{equation}
\label{cusp} r_{mb} < r_{cusp} < r_{ms} \equiv {\rm ISCO}.
\end{equation}
Polish doughnuts have been useful in semi-analytic studies of the
astrophysical appearance of super-Eddington accretion \citep[see
e.g.][]{sik-1971, mad-1988, szu-1996} and in analytic calculations 
of small-amplitude oscillations of accretion structures in
connection with QPOs \citep[see e.g.][]{bla-2006}. 
In the same context, numerical studies of their oscillation properties for different angular momentum distributions were first carried out by \citet{rez-2003a, rez-2003b}. 
Moreover, Polish doughnuts are routinely used as convenient starting initial configurations in numerical simulations \citep[e.g.][]{haw-2001,dev-2003}. Recently, \citet{kom-2006} has constructed analytic models of magnetized 
Polish doughnuts.
However, a closer inspection of Figure \ref{analytic-numerical}
reveals that the numerically constructed model of accretion has a
(much) larger ``vertical'' pressure gradient than the analytic
Polish doughnut, and that in the numerical model the gradient is
visibly enhanced along roughly conical surfaces, approximately
$30^{\circ}$ from the equatorial plane. This (and several other)
detailed features of the accretion structure cannot be modeled by
either the Keplerian nor the constant angular momentum assumption
alone. We suggest and discuss in this paper a simple but flexible
ansatz, that is a combination of the two standard distributions,
Keplerian (\ref{Keplerian}) and constant
(\ref{constant-momentum}). The new ansatz preserves the virtues of
assuming the standard distributions where this is appropriate, but
leads to a far richer variety of possible accretion structures, as
are indeed seen in numerical simulations.

\section{Assumptions and definitions}


We assume that the accretion flow is stationary and axially
symmetric. This assumption expressed in terms of the
Boyer-Lindquist spherical coordinates states that the flow
properties depend only on the radial and polar coordinates $r,
\theta$, and are independent on time $t$ and azimuth $\phi$. We
also assume that the dynamical timescale is much shorter than the
thermal and viscus ones (\ref{timescales}). Accordingly, we ignore
dissipation and assume the stress-energy tensor in the perfect
fluid form,
\begin{equation}
\label{perfect-fluid} 
T^i_{~k} = (p + \epsilon)u^i\,u_k - p\delta^i_{~k},
\end{equation}
with $p$ and $\epsilon$ being the pressure and energy density,
respectively. The four velocity of matter $u^i$ is assumed to be 
purely circular,
\begin{equation}
\label{circular-orbits} 
u^i = (u^t, u^{\phi}, 0, 0).
\end{equation}
The last assumption is not fulfilled close to the cusp (see Figure
\ref{analytic-numerical}), where there is a transition from
``almost circular'' to almost ``free-fall'' radial trajectories.
Nevertheless, the transition could be incorporated in the form of the inner
boundary condition \citep[the relativistic Roche lobe overflow,
see e.g.][]{abr-1985}.

One introduces the specific angular momentum ${\cal L}$, the
angular velocity $\Omega$, and the redshift factor $A$ by the well
known and standard definitions,
\begin{equation}
\label{definitions} {\cal L} = - \frac{u_{\phi}}{u_t}, ~~ \Omega =
\frac{u^{\phi}}{u^t}, ~~ A^{-2} = g_{tt} + 2\Omega g_{t\phi} +
\Omega^2 g_{\phi\phi}.
\end{equation}
The specific angular momentum and angular velocity are linked by
\begin{equation}
\label{velocity-momentum} {\cal L} = - \frac{\Omega\,g_{\phi\phi}
+ g_{t\phi}}{\Omega\,g_{t\phi} + g_{tt}}, ~~ \Omega = -
\frac{{\cal L}\,g_{tt} + g_{t\phi}}{{\cal L}\,g_{t\phi} +
g_{\phi\phi}}.
\end{equation}
The conserved angular momentum $\ell$ is given by,
\begin{equation}
\label{conserved-momentum} \ell = \frac{(p +
\epsilon)u_t}{\rho_0}\,{\cal L}.
\end{equation}


\section{The shapes of the equipressure surfaces}


In this section we briefly discuss one particularly useful result
obtained by \citet{jar-1980}. It states that for a perfect fluid
matter rotating on circular trajectories around a black hole,
the shapes and location of the equipressure surfaces $p(r, \theta)
=~$const follow directly from the assumed angular momentum
distribution (\ref{momentum-distribution}) alone. In particular,
they are independent of the equation of state, $p = p(\epsilon,
{\cal S})$, and the assumed entropy distribution
(\ref{entropy-distribution}).

For a perfect-fluid matter,  the equation of motion
$\nabla_i\,T^i_{~k} = 0$ yields,
\begin{equation}
\label{Euler} \frac{\partial_i p}{p + \epsilon} = -\frac{1}{2}
\frac{\partial_i\,g^{tt} - 2{\cal L}\,\partial_i g^{t\phi} + {\cal
L}^2\,\partial_i g^{\phi\phi}}{g^{tt} - 2{\cal L}\,g^{t\phi} +
{\cal L}^2\,g^{\phi\phi}},
\end{equation}
which may be transformed into,
\begin{equation}
\label{von-Zeipel} \frac{\partial_i p}{p + \epsilon} = \partial_i
\ln A + \frac{{\cal L}\,\partial_i \Omega}{1 - {\cal L}\,\Omega}
\end{equation}
From the second derivative commutator $\partial_r\partial_{\theta}
- \partial_{\theta}\partial_r$ of the above equation,
\begin{equation}
\label{second-commutator-von-Zeipel} \frac{\partial_r
p\,\partial_{\theta}\epsilon -
\partial_{\theta}p\,\partial_r\epsilon}{(p + \epsilon)^2} =
\frac{\partial_r \Omega\,\partial_{\theta}{\cal L} -
\partial_{\theta} \Omega\,\partial_r{\cal L}}{(1 - {\cal
L}\,\Omega)^2},
\end{equation}
one derives \citep[see e.g.][]{abr-1971} the von Zeipel condition:
$p(r,\theta)\,$$=\,$const surfaces coincide with those of
$\epsilon(r,\theta)\,$$= $const, {\it if and only if} the surfaces
${\cal L}(r,\theta)\,$$=\,$const coincide with those
$\Omega(r,\theta)\,$$=\,$const\footnote{The best known Newtonian
version of the von Zeipel condition states that for a barytropic
fluid $p = p(\epsilon)$, both angular velocity and angular
momentum are constant on cylinders, $\Omega = \Omega(R)$, $\cal L
= \cal L(R)$, with $R = r\sin\theta$ being the distance from the
rotation axis.}. Obviously, the constant angular momentum case
satisfies the von Zeipel condition.

\citet{jar-1980} have also discussed a general, non barytropic
case. They wrote equation (\ref{Euler}) twice, for $i = r$ and $i
= \theta$, and divided the two equations side by side to get
\begin{equation}
\label{master} \frac{\partial_r\,p}{\partial_{\theta}\,p} =
\frac{\partial_r\,g^{tt} - 2{\cal L}\,\partial_r g^{t\phi} + {\cal
L}^2\,\partial_r g^{\phi\phi}}{\partial_{\theta}\,g^{tt} - 2{\cal
L}\,\partial_{\theta} g^{t\phi} + {\cal L}^2\,\partial_{\theta}
g^{\phi\phi}} \equiv - F\left(r, \theta \right).
\end{equation}
For the Kerr metric components one knows the functions $g^{ik} =
g^{ik}(r,\theta)$, and therefore the function $F(r,\theta)$ in the
right hand side of (\ref{master}) is known explicitly in terms of
$r$ and $\theta$, {\it if} one knows or assumes the angular
momentum distribution ${\cal L} = {\cal L}(r, \theta)$. This has
an important practical consequence.

Let $\theta = \theta(r)$ be the explicit equation for the
equipressure surface $p(r, \theta) =$const. It is, $d\theta/dr =
-{\partial_r}p/\partial_{\theta}p$. If the function $F(r, \theta)$
in (\ref{master}) is known, then equation (\ref{master}) takes the
form of an ordinary differential equation for the equipressure
surface, $\theta = \theta(r)$,
\begin{equation}
\label{differential} \frac{d\theta}{dr} = F(r, \theta),
\end{equation}
with the explicitly known right hand side. It may be therefore
directly integrated to get all the possible locations for the
equipressure surfaces.


\section{The angular momentum distribution}
\label{section-angular-momentum}


\subsection{Physical arguments: the radial distribution}
\label{section-physical-arguments}

\citet{jar-1980} discussed general arguments showing that the
slope of the specific angular momentum should be between two
extreme: the slope corresponding to ${\cal L} =\,$const and the
slope corresponding to $\Omega =\,$const. These two cases,
together with the Keplerian one ${\cal L} = {\cal L}_K$, may be
considered as useful archetypes in discussing arguments relevant
to the angular momentum distribution.

Indeed, far away from the black hole $r \gg r_G$, these arguments
are well known \cite[see e.g.][]{fra-2002} and together with
numerous numerical simulations show that typically (i.e. in a
stationary case with no shocks) the specific angular momentum
should be slightly sub-Keplerian ${\cal L}(r, \pi/2) \approx {\cal
L}_K(r)$. There is a solid consensus on this point.

The situation close to the black hole is less clear because there is
not sufficient knowledge of the nature of the stress operating in
the innermost part of the flow, i.e. approximately between the
horizon and the ISCO. Formally, one may consider two extreme ideal
cases, depending whether the stress is very small or very large.

In the first case, the almost vanishing stress implies that the
fluid is almost free-falling, and therefore the angular momentum
is almost constant along fluid lines. This leads to ${\cal L}(r,
\pi/2) \approx\,$const. Such situation is typical for the thin
\citet{sha-sun-1974} and slim \citep{abr-1988} accretion disks. In
the second case, one may imagine a powerful instability like MRI,
which occurs when $d\Omega/dr \ne 0$. It may force the fluid
closer to the marginally stable state $\Omega =\,$const. This
situation may be relevant for ADAFs \citep{nar-1995, abr-1995}.
 \begin{figure*}
  \centering
   \includegraphics[width=4.45cm,height=4.45cm]{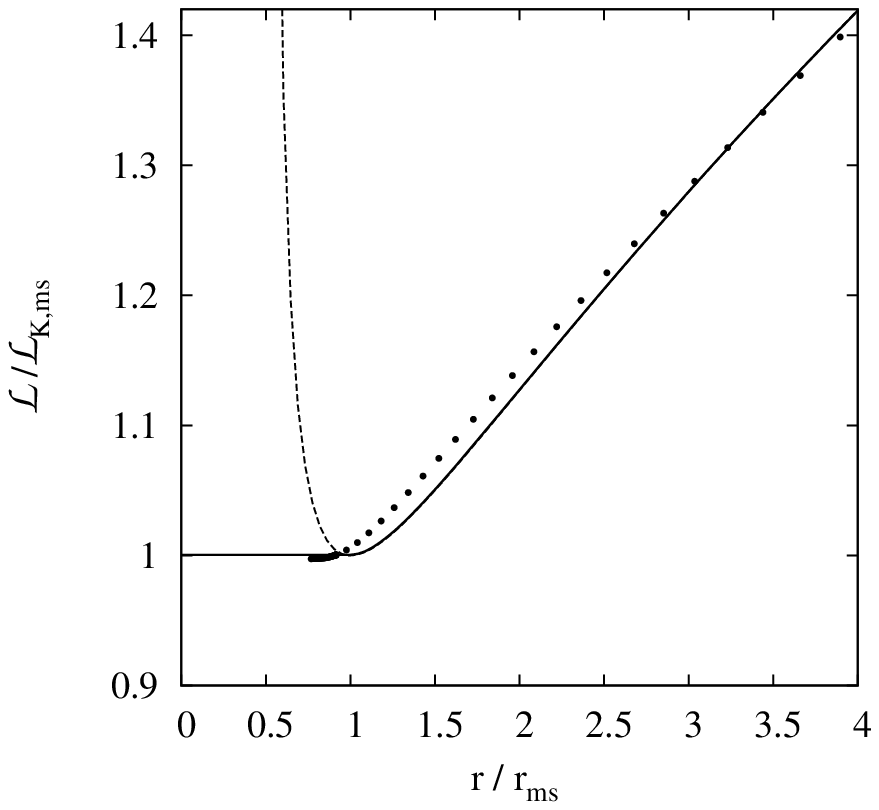}
   \hfill
   \includegraphics[width=4.45cm,height=4.45cm]{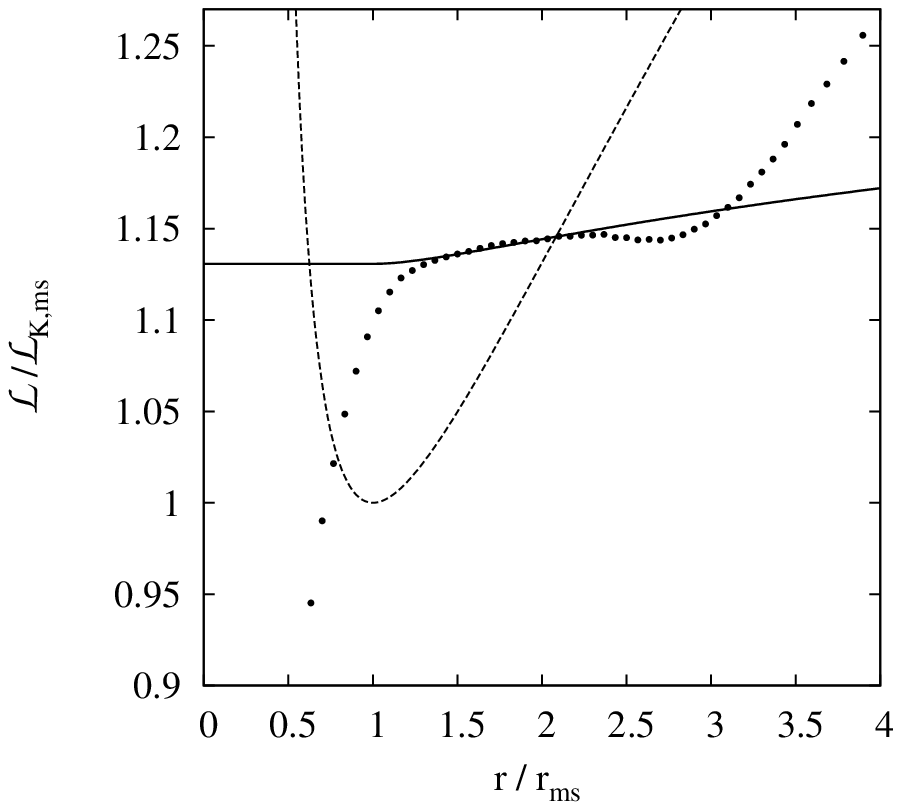}
   \hfill
   \includegraphics[width=4.45cm,height=4.45cm]{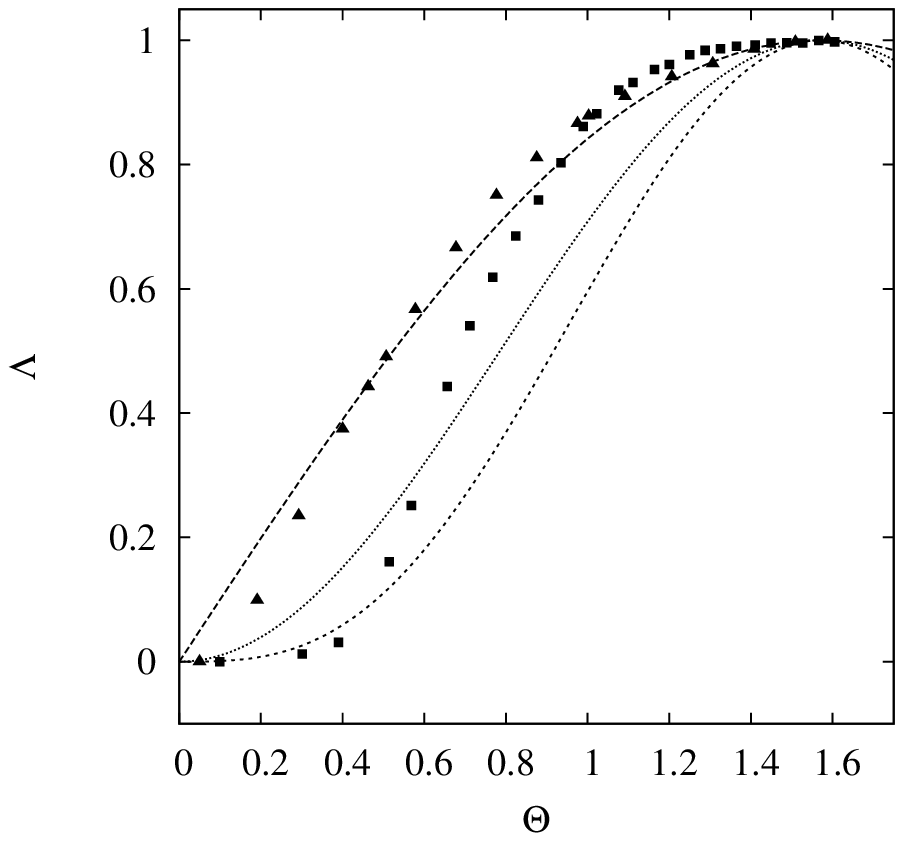}
   \hfill
   \includegraphics[width=4.45cm,height=4.45cm]{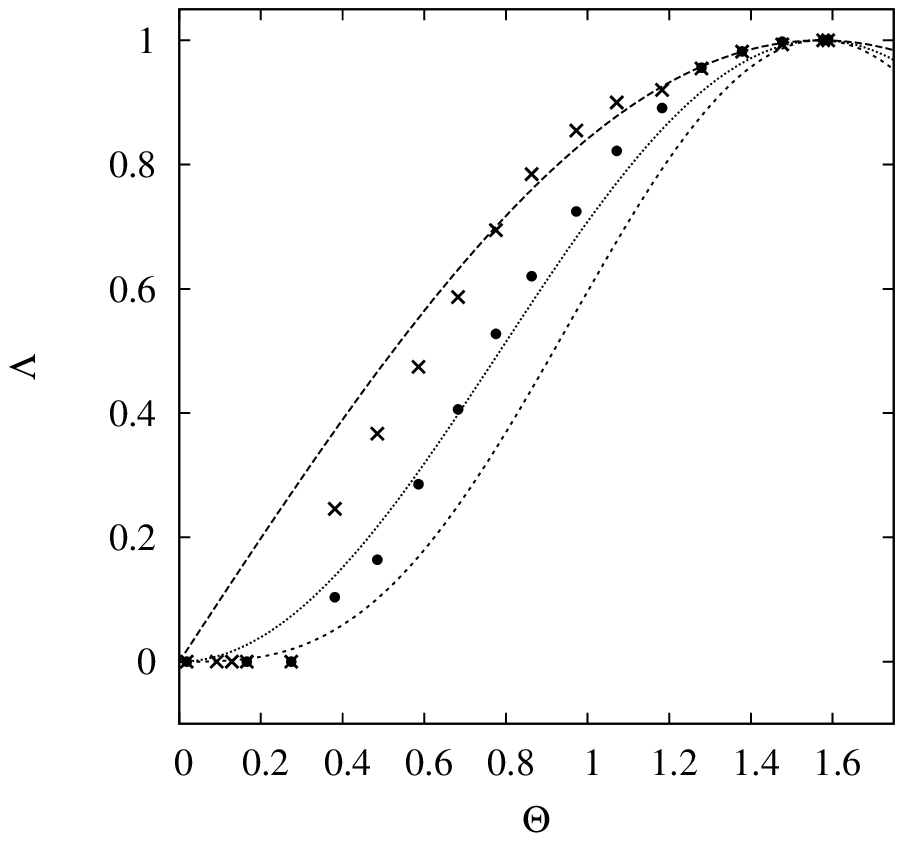}
      \caption{
      (a) and (b): the distribution of angular momentum on the equatorial plane. Thick lines correspond to the
      angular momentum predicted by our analytic formula, dashed lines show the Keplerian angular momentum distribution and dots to the simulation data. (a): Kerr geometry $a=0.9$ simulations by \citet{sad-2008}. (b): Pseudo-Newtonian MHD simulations by \citet{mac-2008}. (c) and (d): angular momentum off the equatorial plane, normalized to its equatorial plane value, $\Lambda = {\cal L}(r,\theta)/{\cal L}(r, \pi/2)$. Lines correspond to the $\sin^{2\gamma}\theta$ distribution at $r = 10r_G$: long-dashed $\gamma=0.5$, dotted $\gamma = 1.0$, and short dashed $\gamma = 1.5$. Points are taken from time-dependent, fully 3-D, MHD numerical simulations. They correspond to time and azimuthal averages at the same radial location, $r = 10r_G$. (c): Points from the simulations of \citet{mac-2008} in the Paczy{\'n}ski-Wiita potential --- triangle: High temperature case; square: Low temperature case. (d): Points from the simulations of \citet{fra-2007} in Schwarzschild (dots) and $a=0.9$ Kerr (crosses) spacetimes.
      }
     \label{fig:ang-mom}
   \end{figure*}

\subsection{The new ansatz}
\label{section-ansatz}

We suggest adopting the following assumption for the angular
momentum distribution,
\begin{equation}
\label{ansatz-general}
  {\cal L}(r, \theta) = \left\{
  \begin{array}{ll}
  {\cal L}_0\left( \frac{{\cal L}_K
(r)}{{\cal L}_0}\right)^{\beta}
          \sin^{2\gamma}\theta & \mbox{~for~ $r \geq r_{ms}$}\\
          ~\\
         {\cal L}_{ms}(r) \sin^{2\gamma}\theta & \mbox{~for~ $r < r_{ms}$}
         \end{array} \right\}
\end{equation}
The constant ${\cal L}_0$ is defined by ${\cal L}_0 \equiv
\eta\,{\cal L}_K(r_{ms})$. For the ``hydrodynamical'' case, the
function ${\cal L}_{ms}(r)$ is constant,
\begin{equation}
\label{constant-definitions-hydro}
 {\cal L}_{ms}(r) = {\cal L}_0\,[{\cal L}_K(r_{ms})/{\cal L}_0]^{\beta}
 = {\rm const},
\end{equation}
while for the ``MHD'' case its is calculated from the $\Omega(r) =
\Omega_K(r_{ms}) =\,$const condition,
\begin{equation}
\label{constant-definitions-MHD} {\cal L}_{ms}(r) = -
\frac{\Omega_{ms}\,g_{\phi\phi}(r, \pi/2) + g_{t\phi}(r,
\pi/2)}{\Omega_{ms}\,g_{t\phi}(r, \pi/2) + g_{tt}(r, \pi/2)}.
\end{equation}
Thus, there are only {\it three} dimensionless parameters in the
model: ($\beta$, $\gamma$, $\eta$). Their ranges are,
\begin{equation}
\label{constants-range} 0 \le \beta \le 1, ~~-1 \le \gamma \le 1,
~~~~1 \le \eta \le \eta_{max}.
\end{equation}
The function ${\cal L}_K(r)$ is the Keplerian angular momentum in
the equatorial plane, $\theta =\pi/2$, which for the Kerr metric
is described by formula (\ref{Keplerian}) and $\eta_{max} = {\cal
L}_K(r_{mb})/{\cal L}_K(r_{ms})$. An equipressure surface that
starts from the cusp is marginally bound for $\beta = 0$, $\gamma
= 0$ and $\eta = \eta_{max}$.

\subsection{Angular momentum on the equatorial plane}

On the equatorial plane, $\sin \theta = 1$, and therefore only
$\beta$ and $\eta$ (through ${\cal L}_0$) enter the distribution
formulae (\ref{ansatz-general}).
\begin{equation}
\label{ansatz-equatorial}
  {\cal L}(r, \pi/2) = \left\{
  \begin{array}{ll}
  {\cal L}_0\left( \frac{{\cal L}_K
(r)}{{\cal L}_0}\right)^{\beta}
          & \mbox{~for~ $r \geq r_{ms}$}\\ ~\\
         {\cal L}_{ms} & \mbox{~for~ $r < r_{ms}$}
         \end{array} \right\}
\end{equation}
When $\beta = 0$, the  angular momentum is constant, ${\cal L} =
{\cal L}_0$, and when $\beta = 1$, it equals the Keplerian one,
${\cal L} = {\cal L}_K$.

For small values of $\beta$ the assumed equatorial plane angular
momentum (\ref{ansatz-equatorial}) reproduces the characteristic
shape, shown in Figure~\ref{fig:ang-mom}, which has been
found in many numerical simulations of accretion flows ---
including stationary, axially symmetric, $\alpha$ viscosity,
hydrodynamical ``slim disks'' \citep[e.g.][]{abr-1988}, and more
recent, fully 3-D, non-stationary MHD simulations
\citep[e.g.][]{mac-2008, fra-2008}. It corresponds to a
distribution that is slightly sub-Keplerian for large radii, and
closer to the black hole it crosses the Keplerian distribution
twice, at $r_{center} > r_{ms}$ and at $r_{cusp} < r_{ms}$,
forming a super-Keplerian part around $r_{ms}$. For $r < r_{cusp}$
the angular momentum is almost constant.

\subsection{Angular momentum off the equatorial plane}

Numerical simulations show that away from the equatorial plane, the angular momentum falls off.
Figure~\ref{fig:ang-mom} shows that indeed several MHD simulations \citep[][Figure 2c and 2d respectively]{mac-2008, fra-2008}, feature a drop of angular momentum away from the equatorial plane.
This behavior is reflected by the term $\sin^{2\gamma}\theta$ in (\ref{ansatz-general}). One may see that this form accurately mimics the outcome of the numerical simulations.
\citet{pro-2003a,pro-2003b} also studied axisymmetric accretion
flows with low specific angular momentum using numerical
simulations. In their inviscid hydrodynamical case
\citet{pro-2003a} found that the inner accretion flow
settles into a pressure-rotation supported torus in the equatorial
region and a nearly radial inflow in the polar funnels.
Furthermore, the specific angular momentum in the equatorial torus
was nearly constant. This behavior changes once magnetic fields
are introduced, as shown in \citet{pro-2003b}. In the MHD case,
the magnetic fields transport specific angular momentum so that in
the innermost part of the flow, rotation is sub-Keplerian, whereas
in the outer part, it is nearly Keplerian. Similar rotational
profiles are also found in MHD simulations of the collapsar model
of gamma-ray bursts \citep{pro-2003c, bai-2008}, which use a sophisticated
equation of state and neutrino cooling (instead of a simple
adiabatic equation of state). Therefore, it appears that the
rotational profile assumed in our model is quite robust as it has
been obtained in a number of numerical experiments with various
microphysics.


\section{Results}

\begin{figure*}
  \centering
   \includegraphics[width=4.3cm]{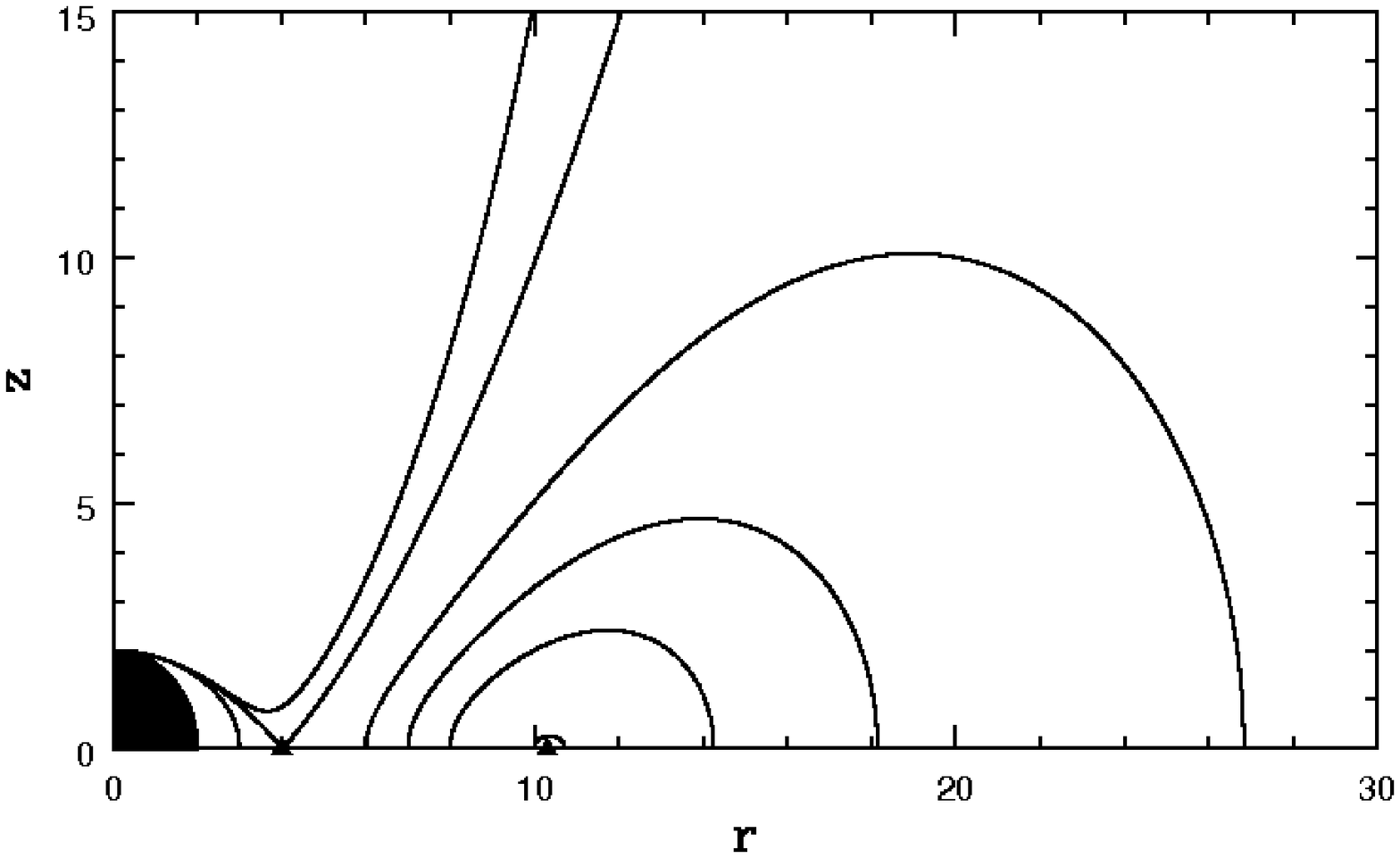}
   \includegraphics[width=4.3cm]{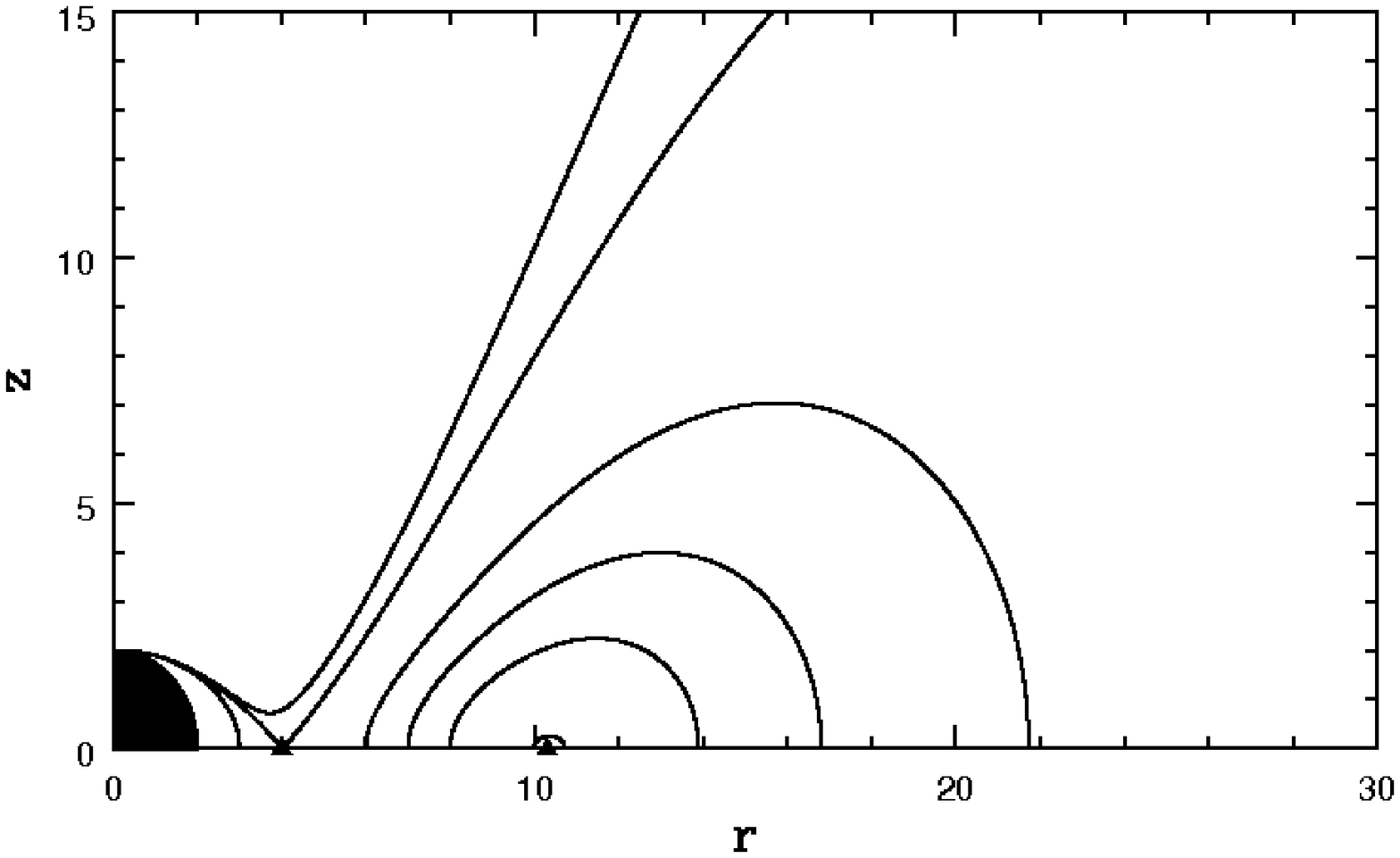}
   \includegraphics[width=4.3cm]{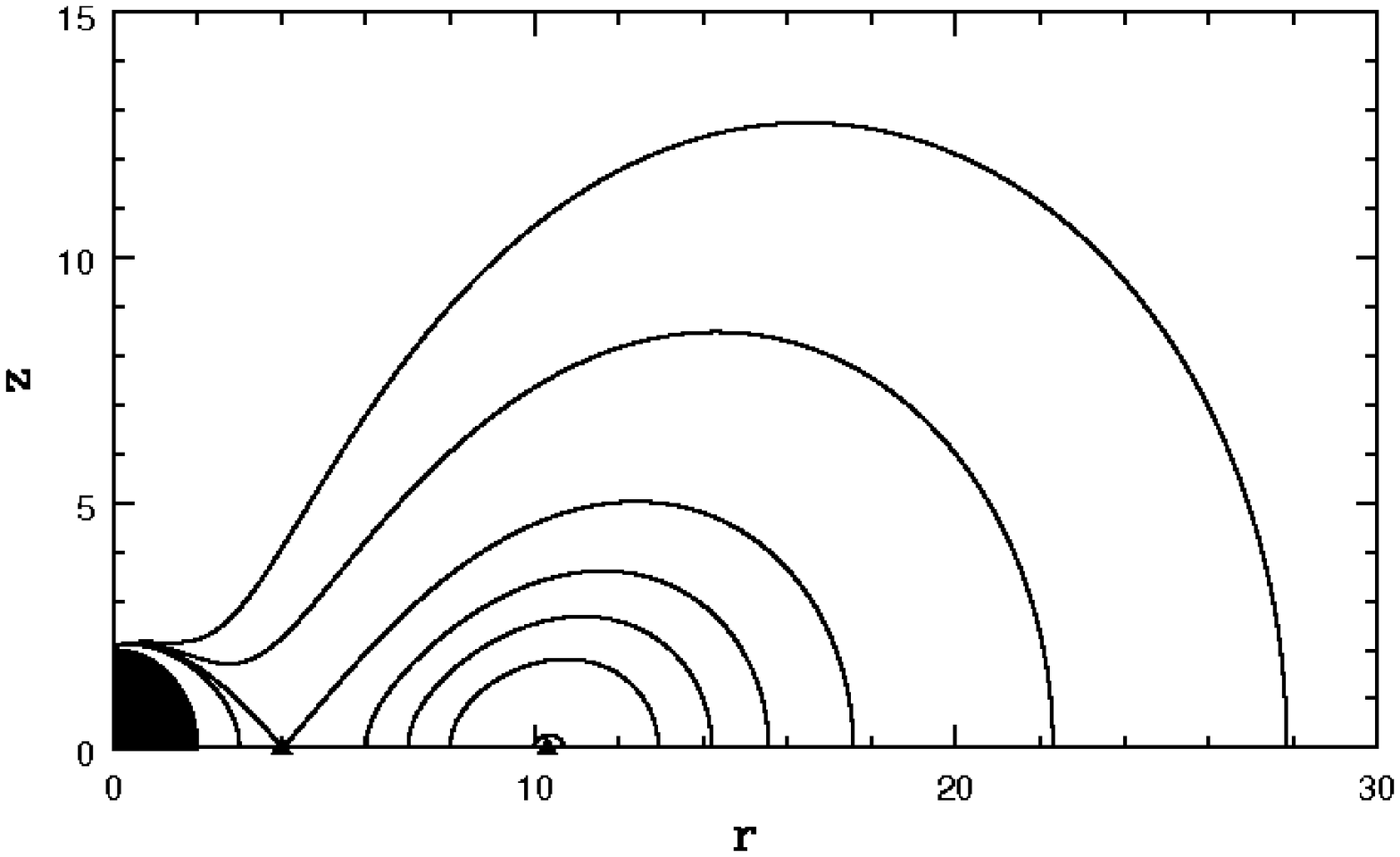}
   \includegraphics[width=4.3cm]{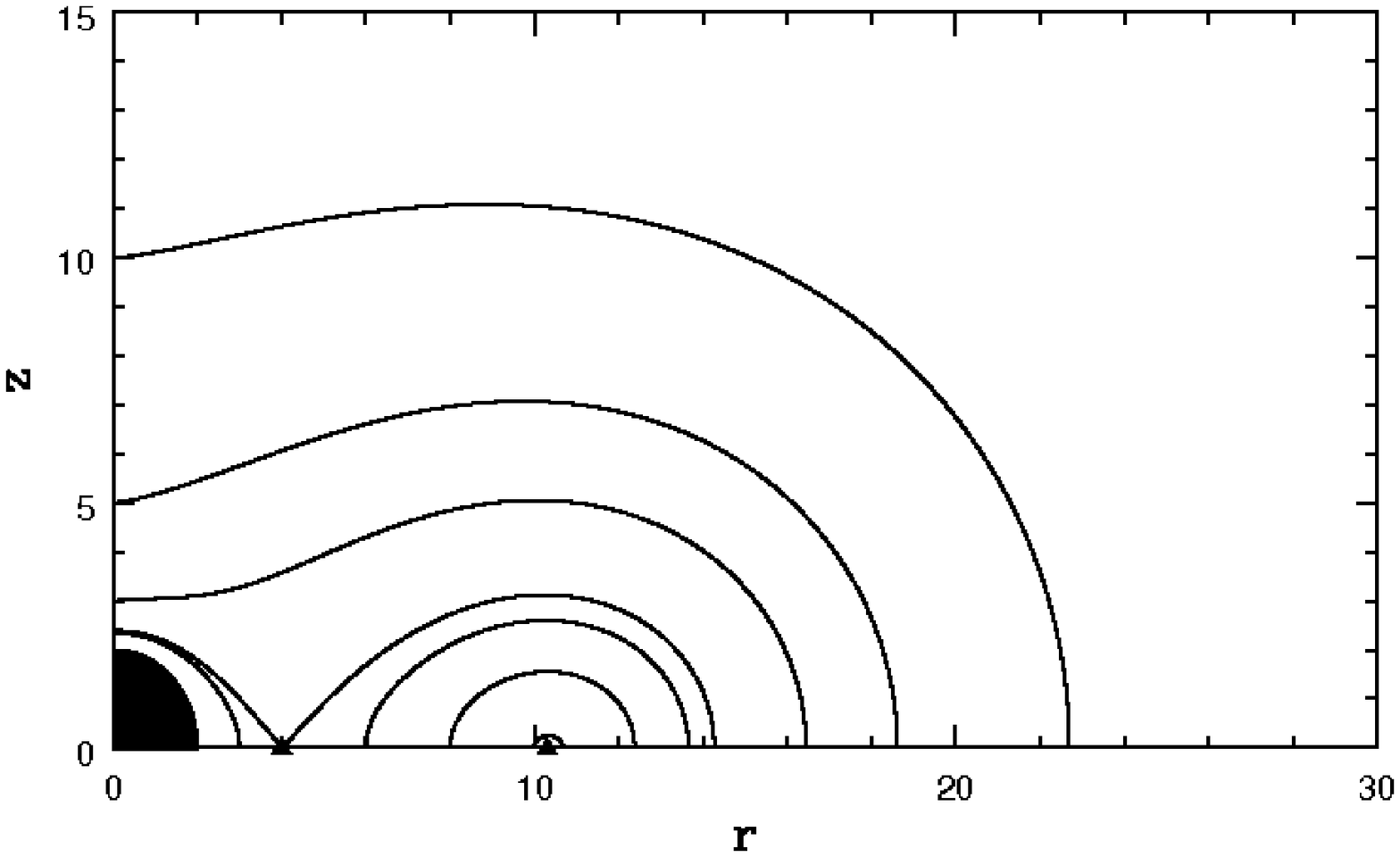}
   \includegraphics[width=4.3cm]{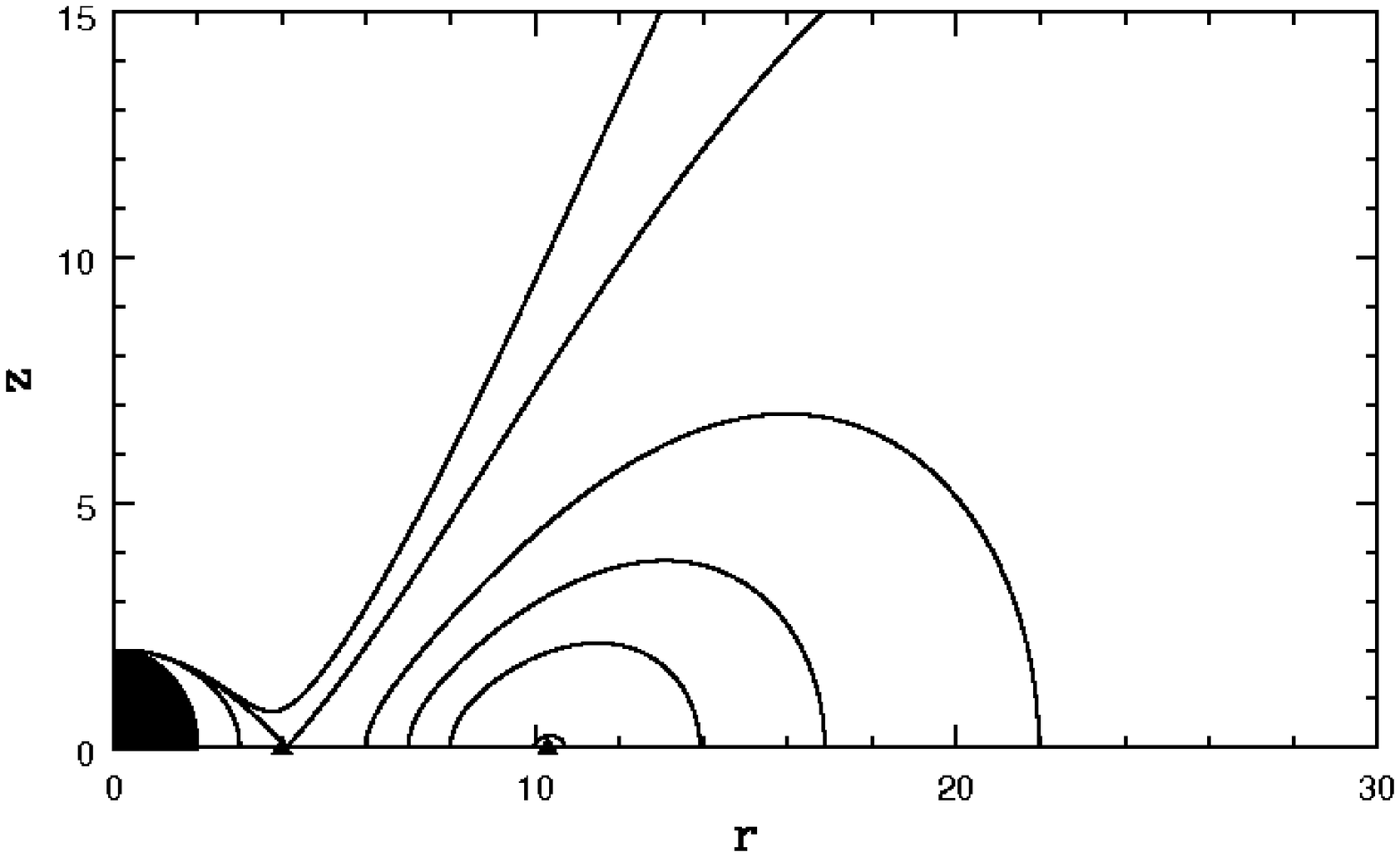}
   \includegraphics[width=4.3cm]{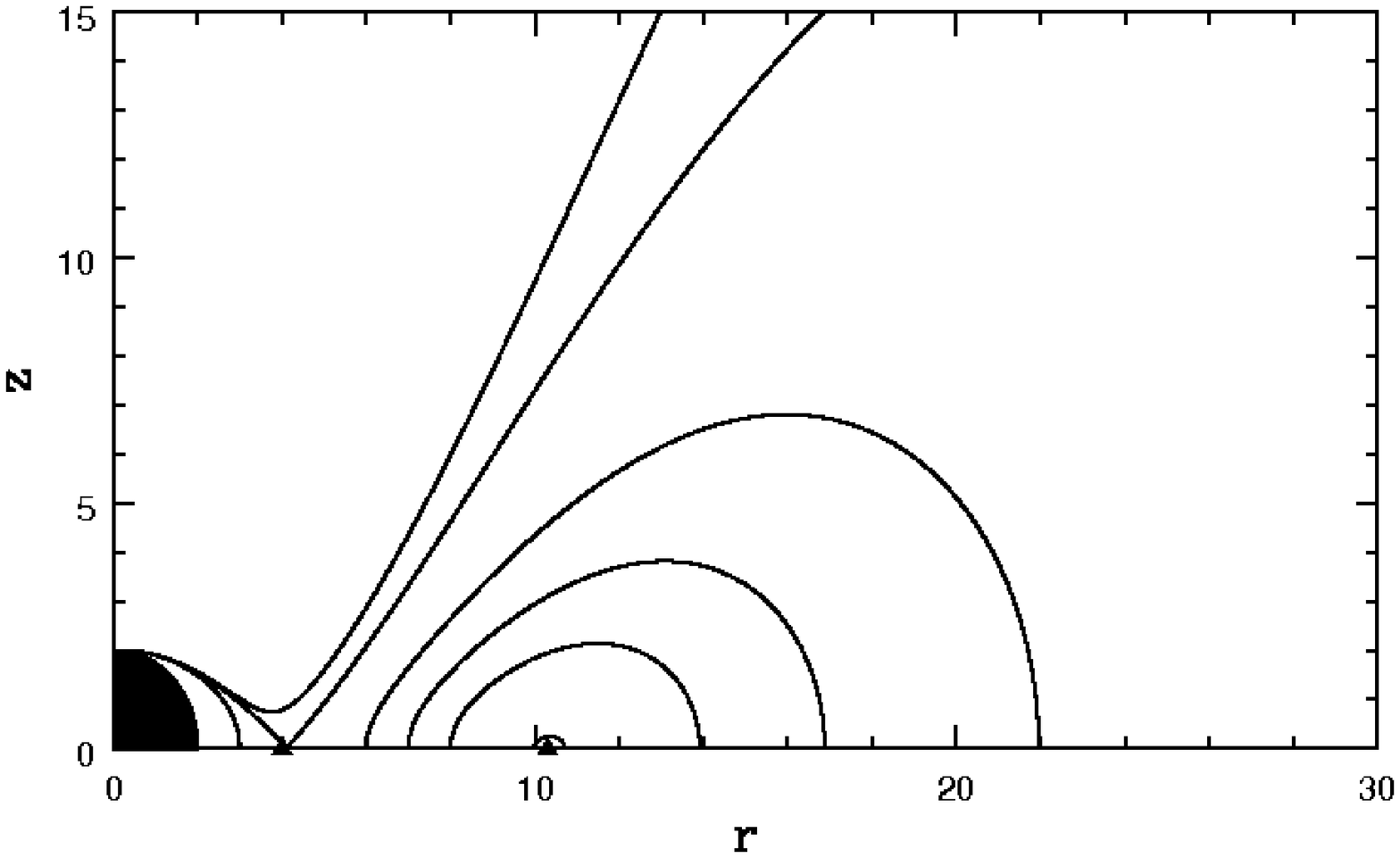}
   \includegraphics[width=4.3cm]{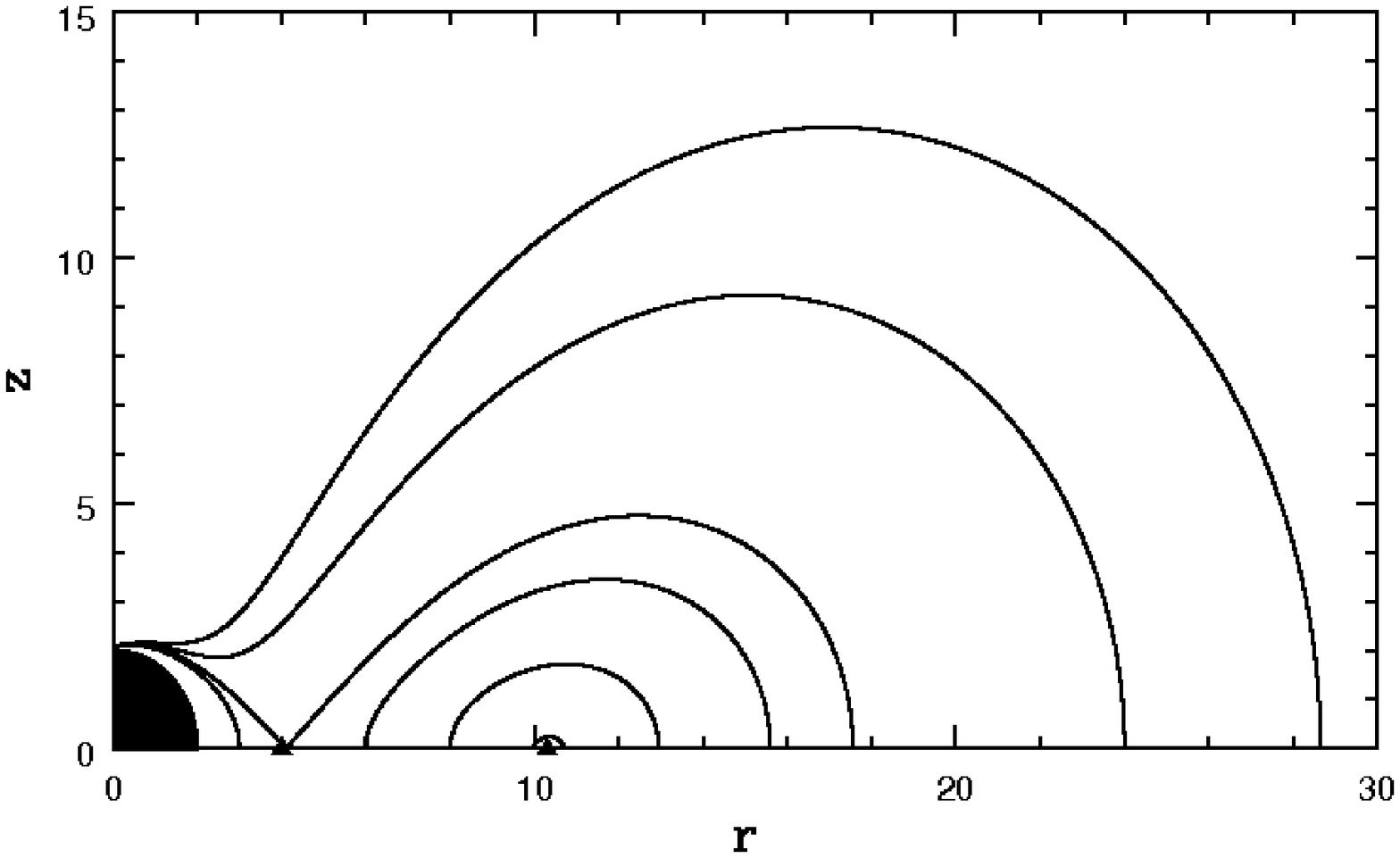}
   \includegraphics[width=4.3cm]{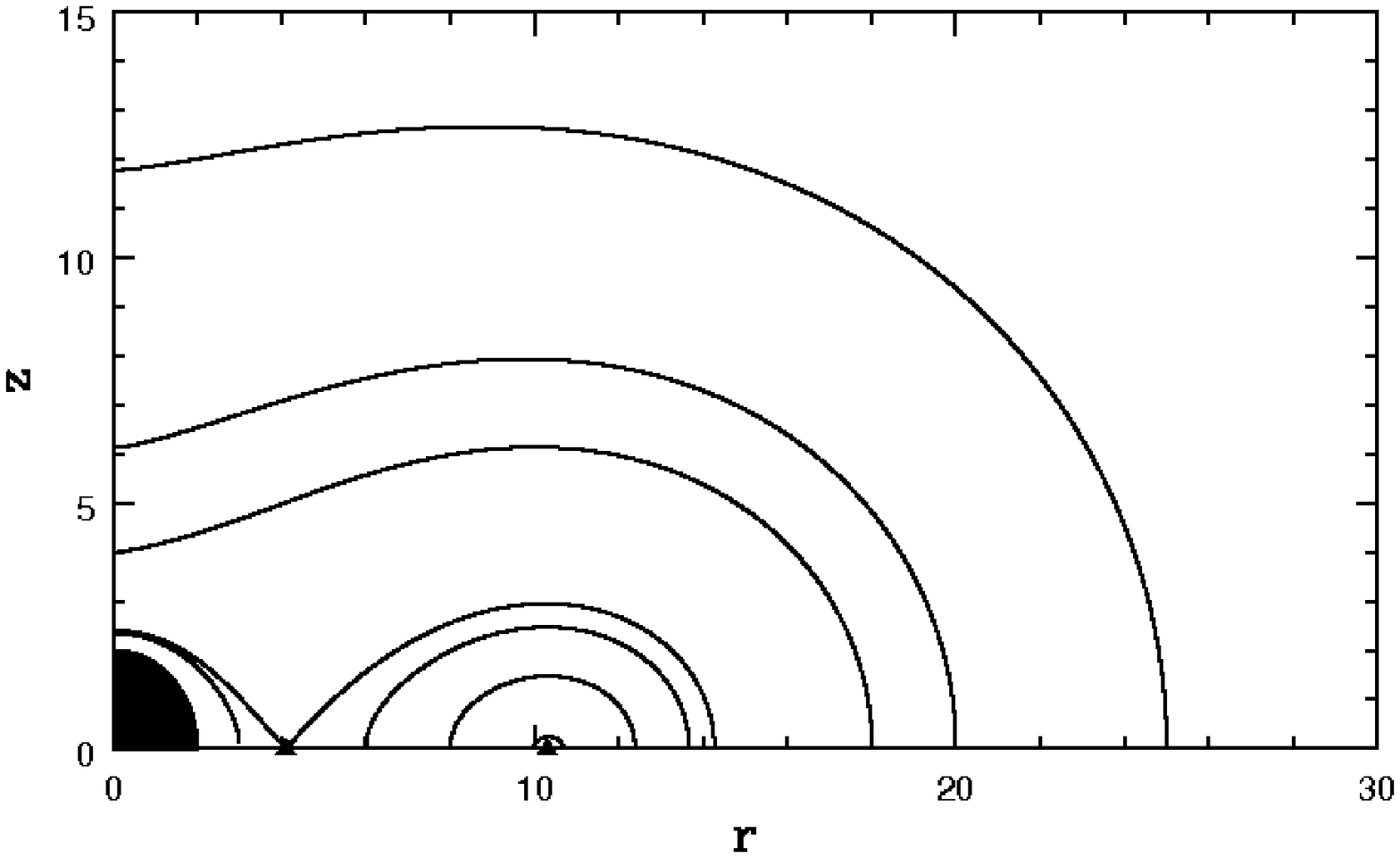}
   \includegraphics[width=4.3cm]{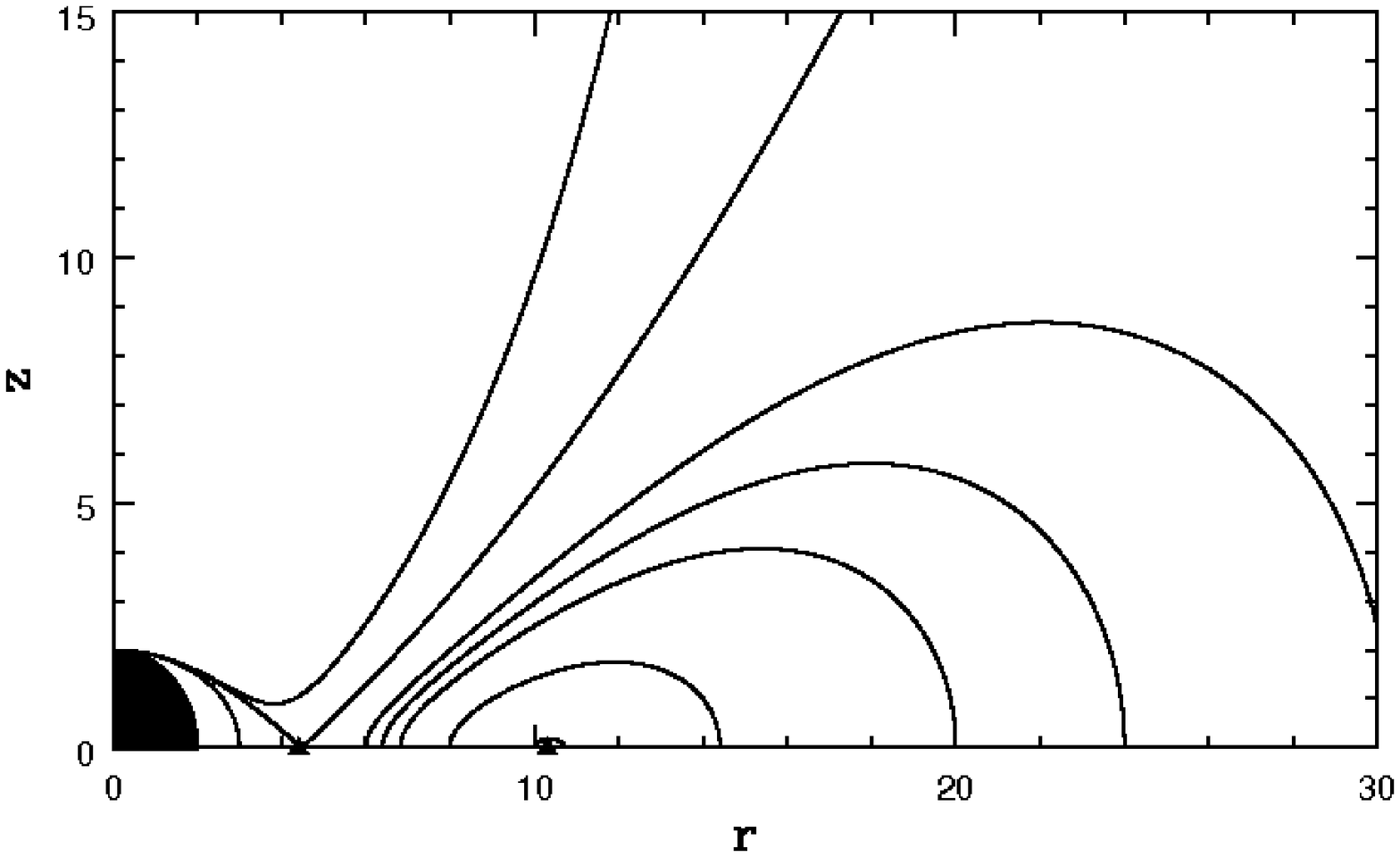}
   \includegraphics[width=4.3cm]{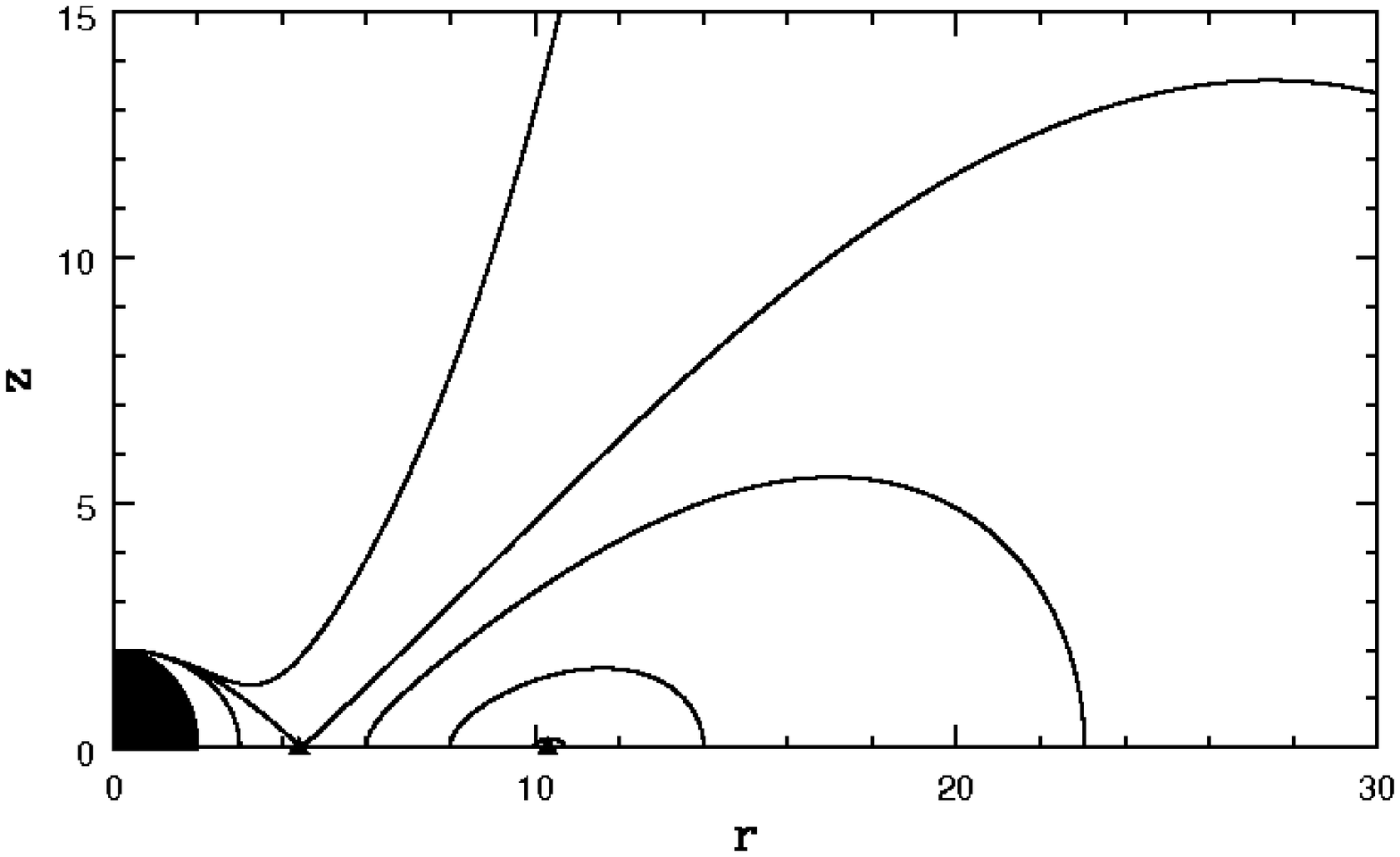}
   \includegraphics[width=4.3cm]{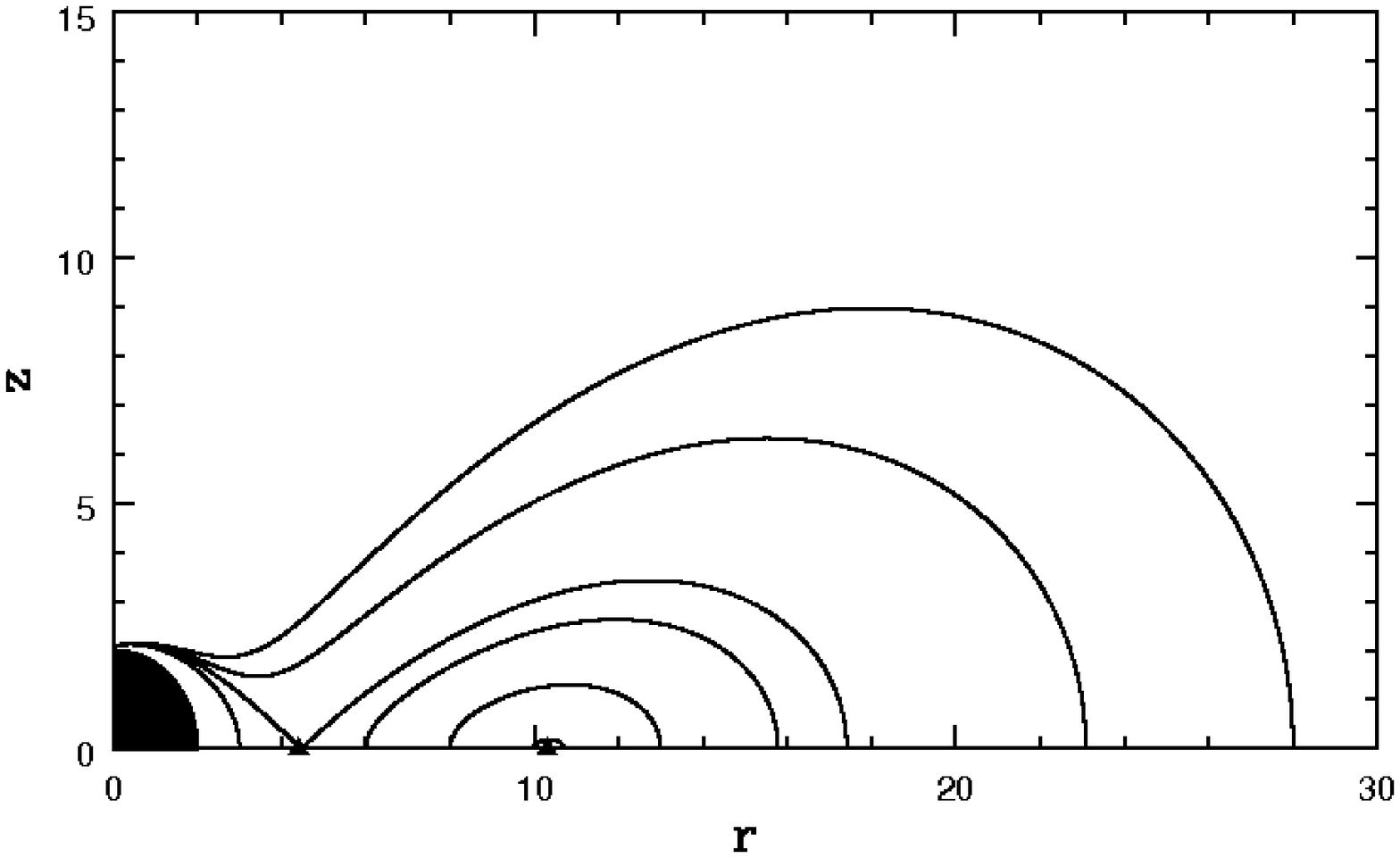}
   \includegraphics[width=4.3cm]{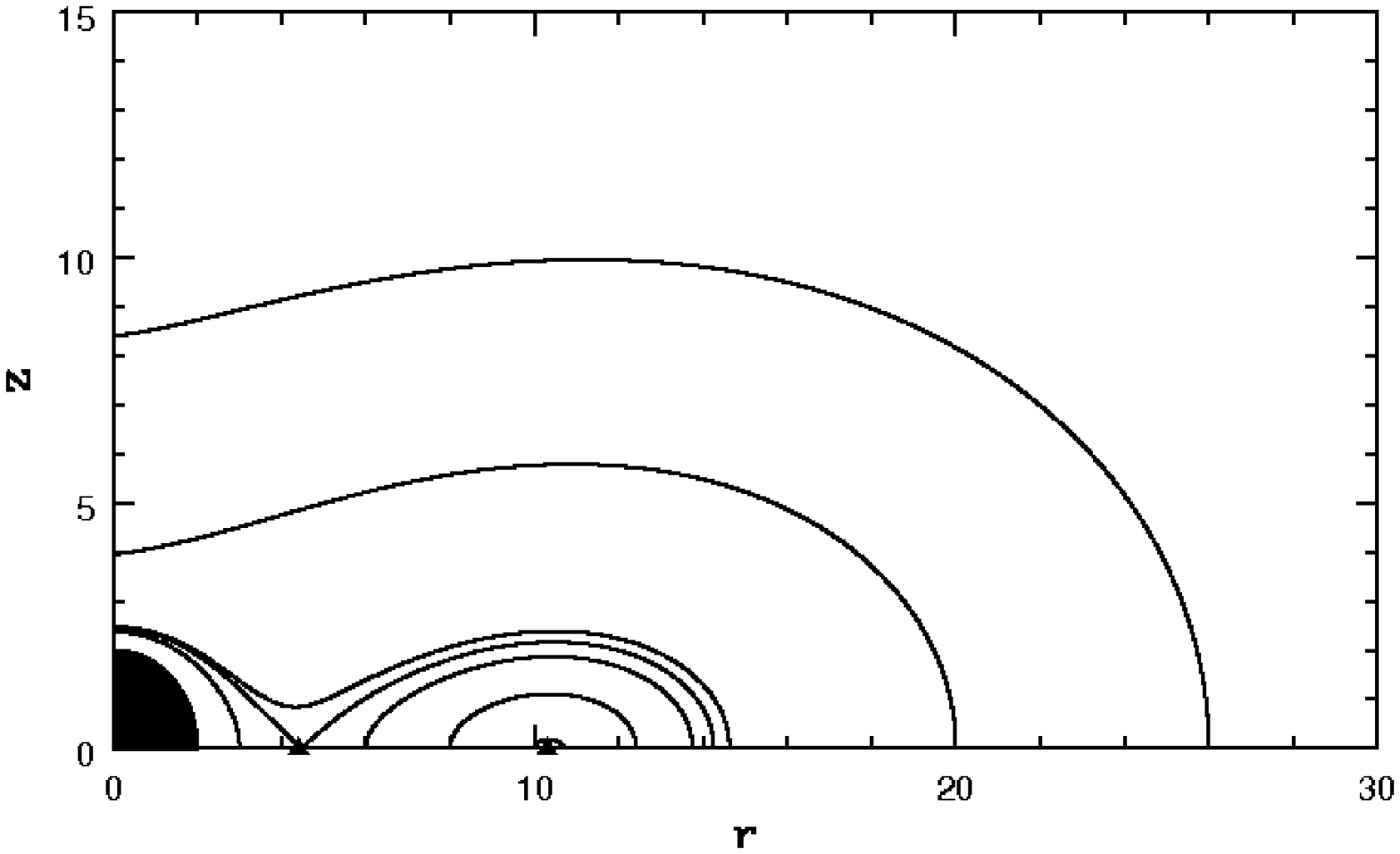}
   \includegraphics[width=4.3cm]{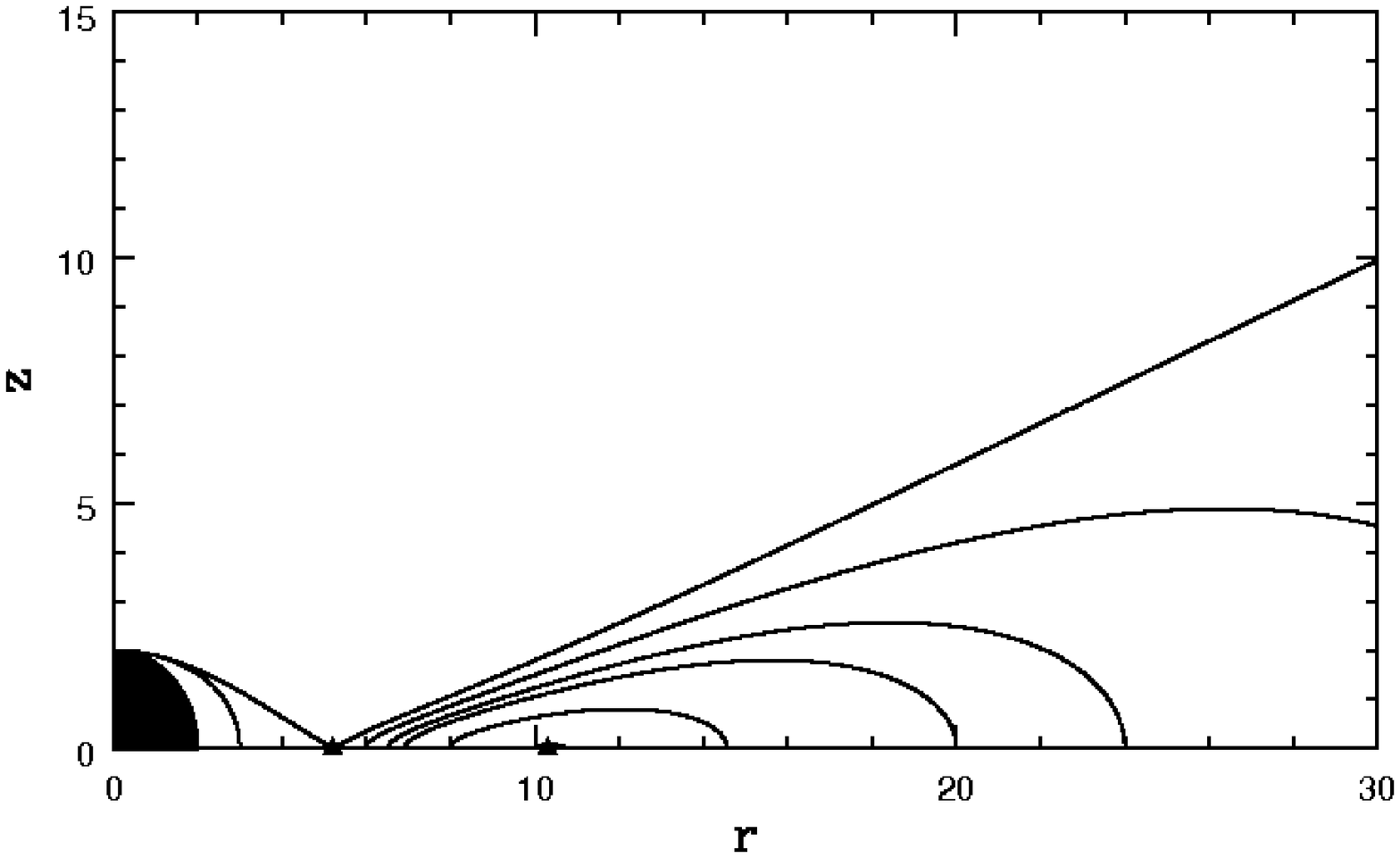}
   \includegraphics[width=4.3cm]{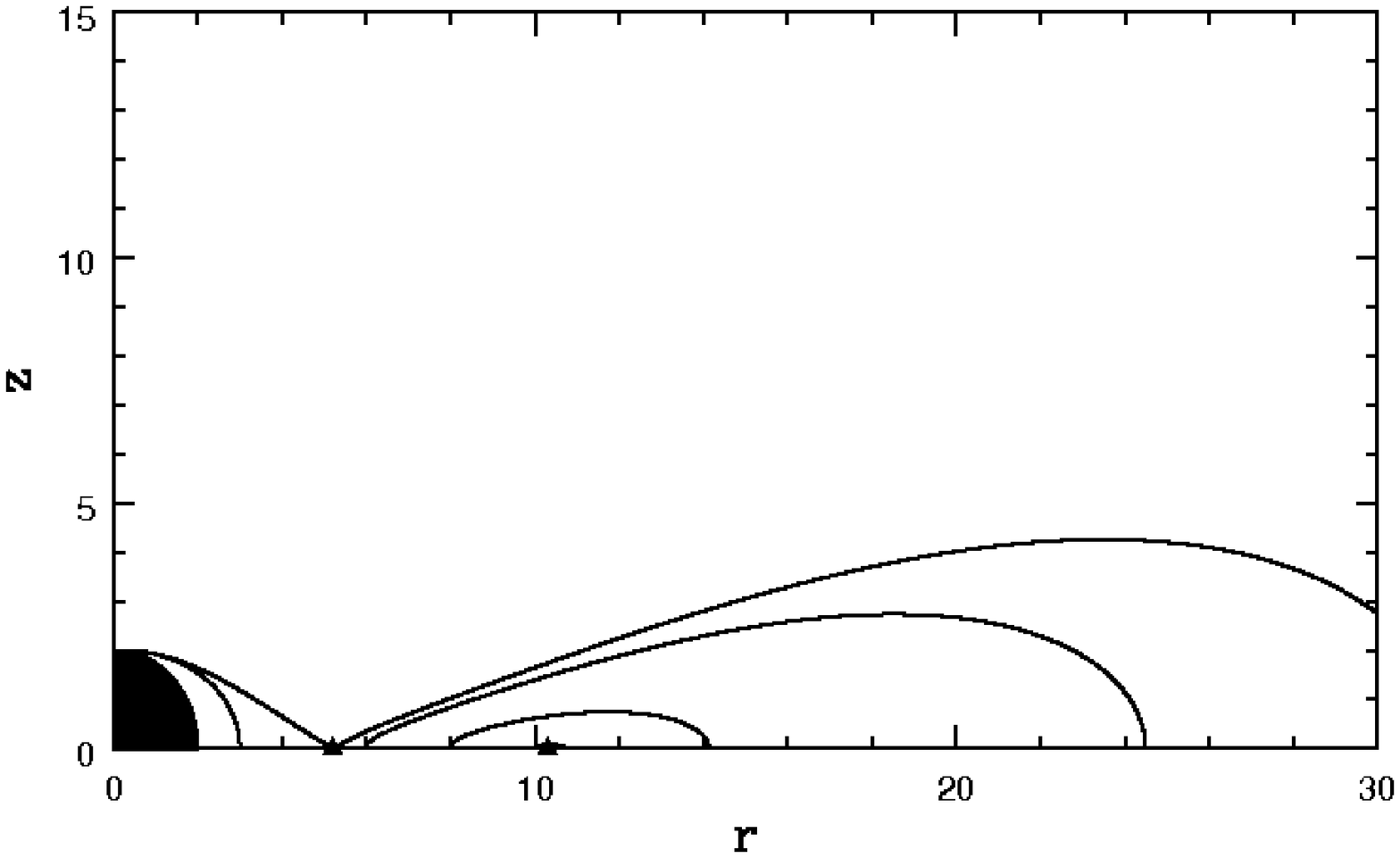}
   \includegraphics[width=4.3cm]{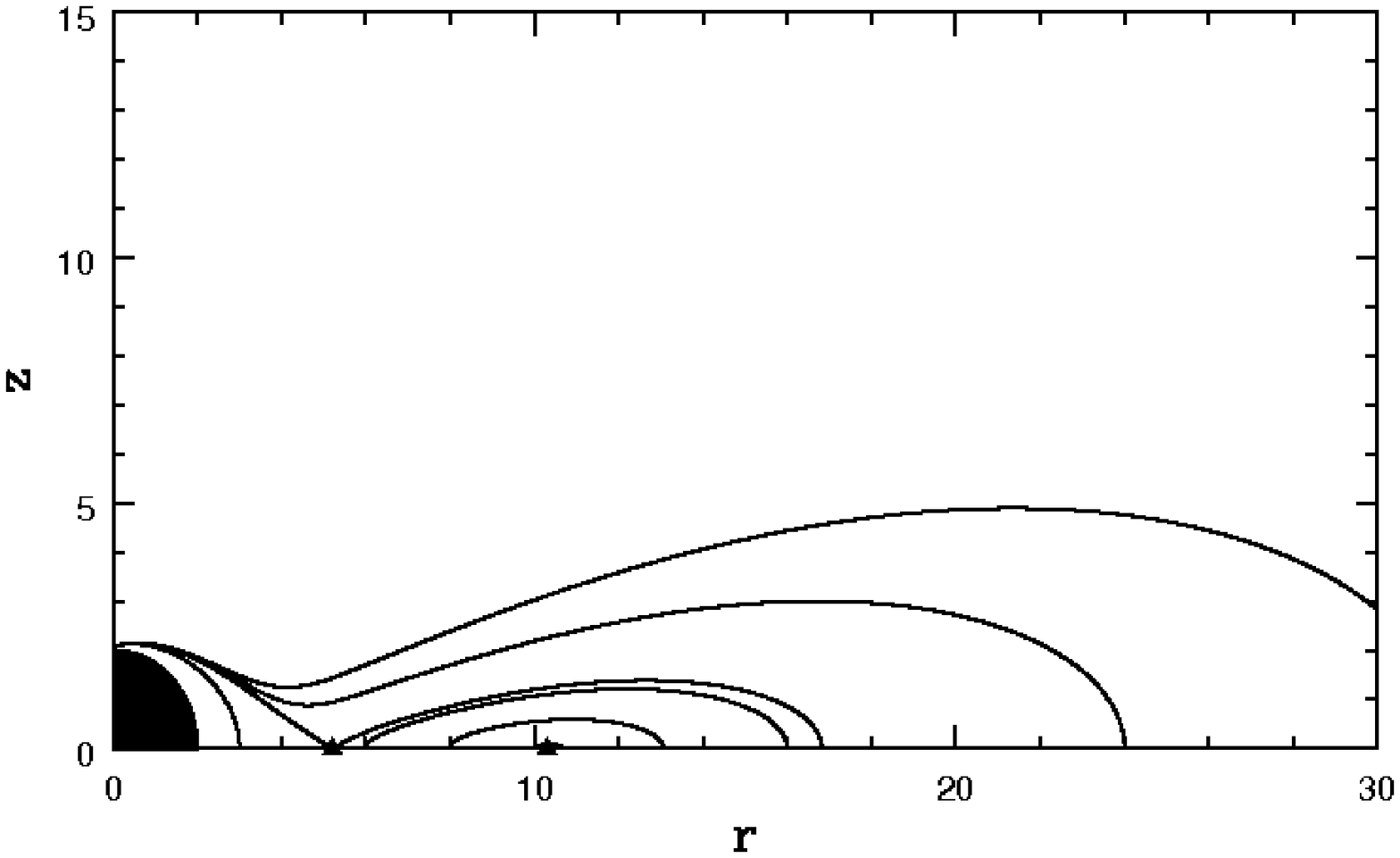}
   \includegraphics[width=4.3cm]{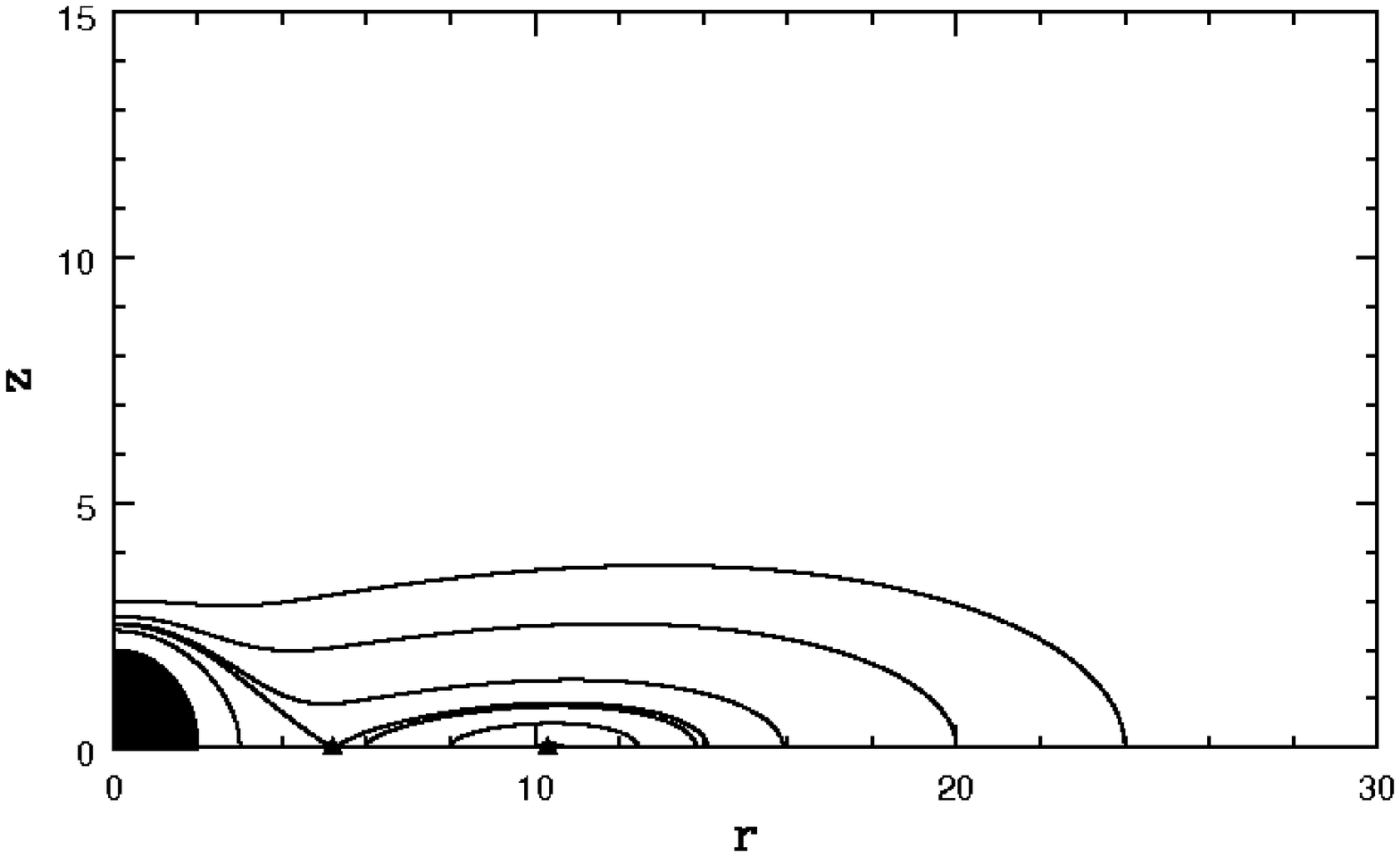}
   \includegraphics[width=4.3cm]{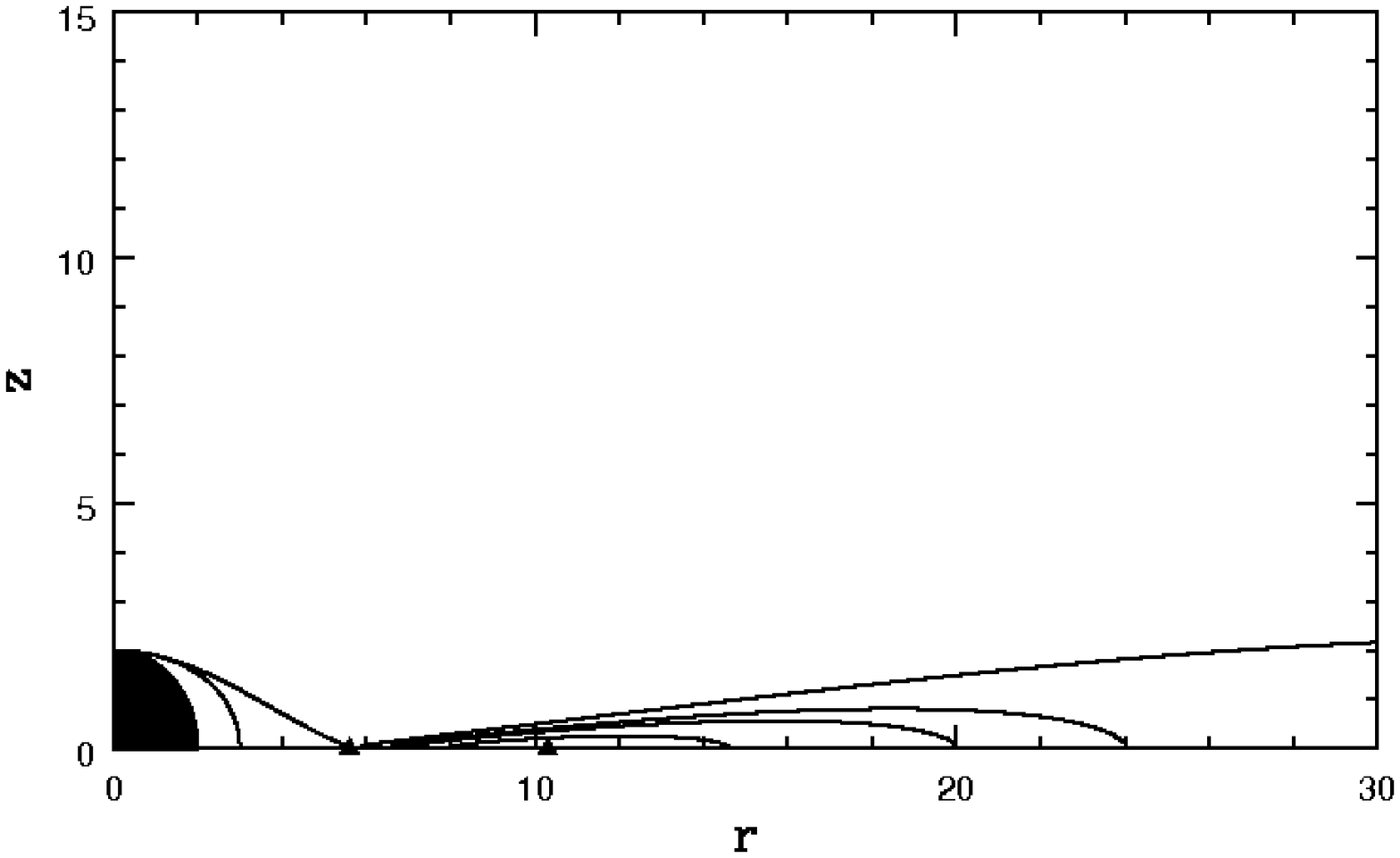}
   \includegraphics[width=4.3cm]{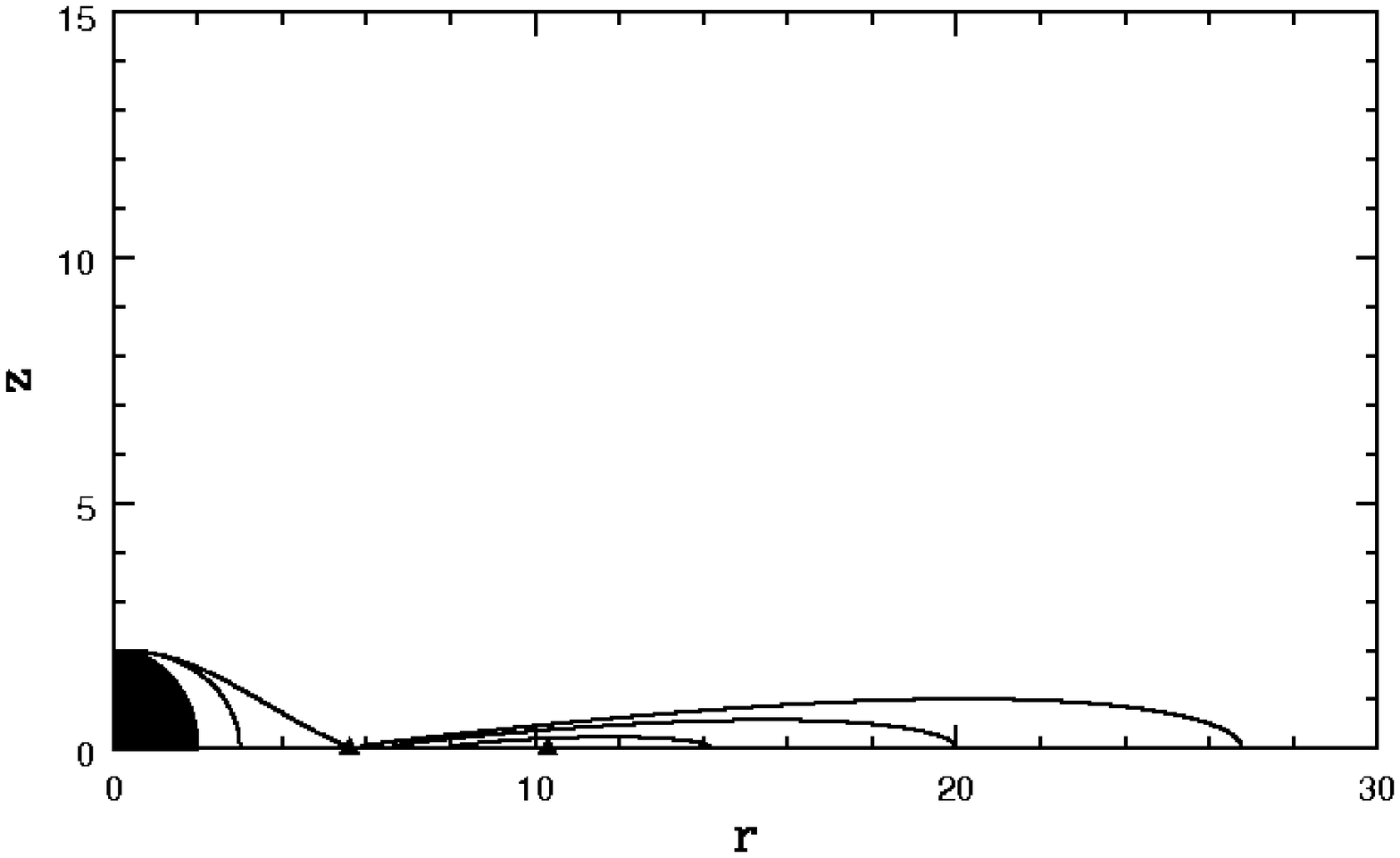}
   \includegraphics[width=4.3cm]{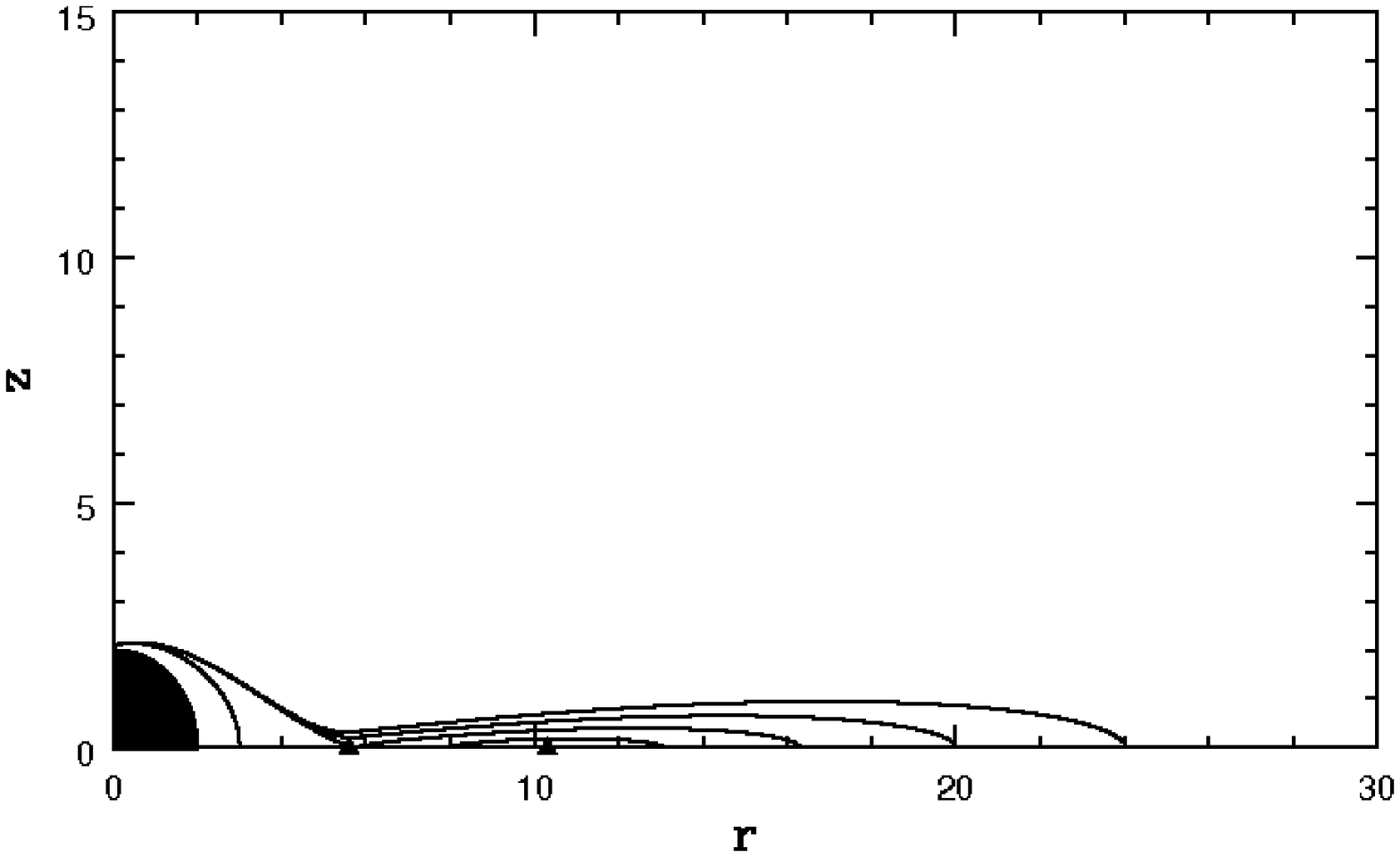}
   \includegraphics[width=4.3cm]{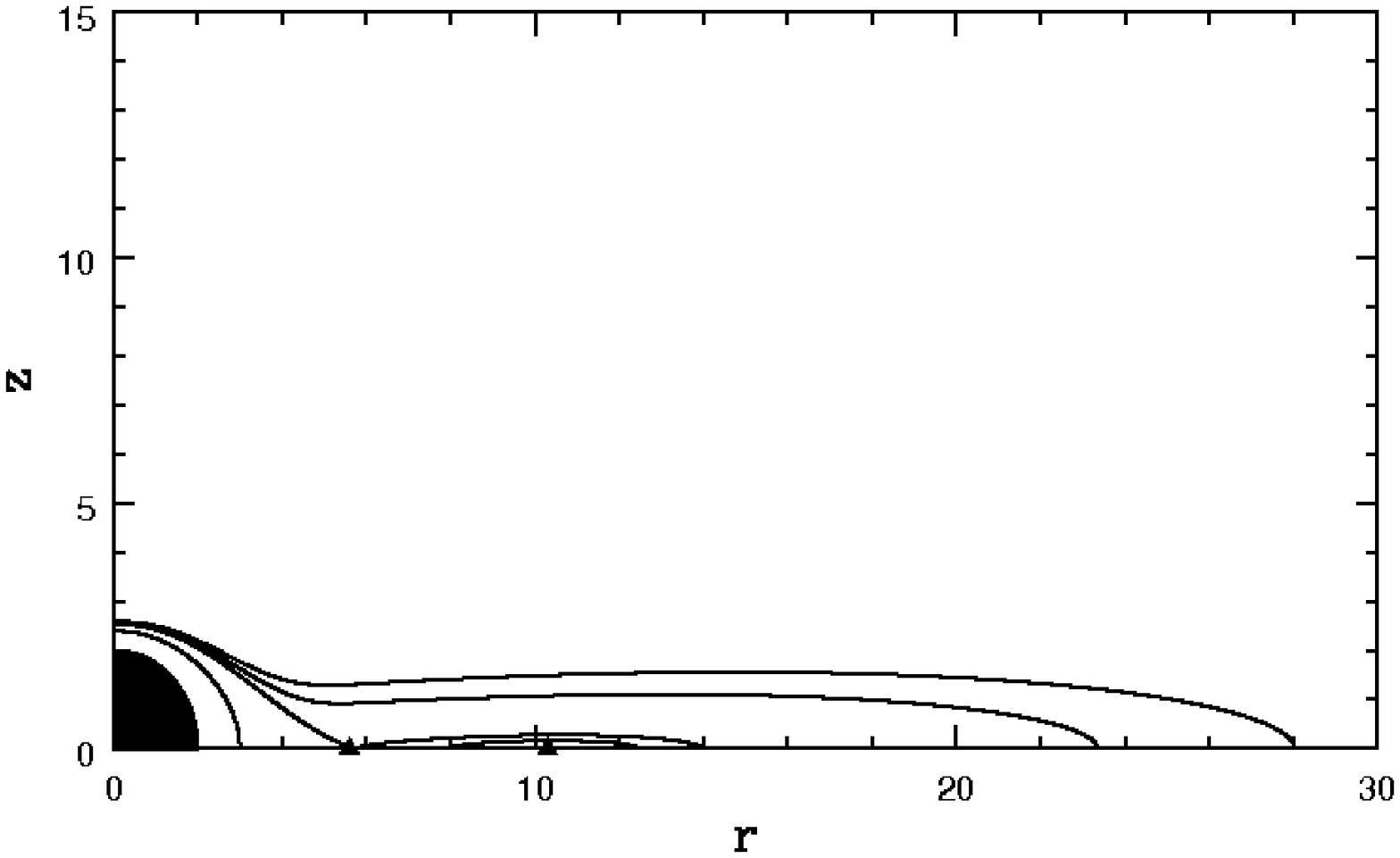}
      \caption
      {Equipressure surfaces for $a = 0$ and $\eta = \eta_{max} = 1.085$.
      Five rows correspond to $\beta = (0.0), (0.1), (0.5), (0.9), (0.99)$
      from the top to the bottom.
      Four columns correspond to $\gamma = (0.0), (0.1), (0.5), (0.9)$
      from the left to the right.
      The upper left corner shows a ``standard'' Polish doughnut. The
      lower right corner shows an almost Keplerian disk at the
      equatorial plane, surrendered by a very low angular momentum
      envelope.
      }
         \label{sequence-beta-gamma}
   \end{figure*}
\begin{figure*}
  \centering
   \includegraphics[width=4.3cm]{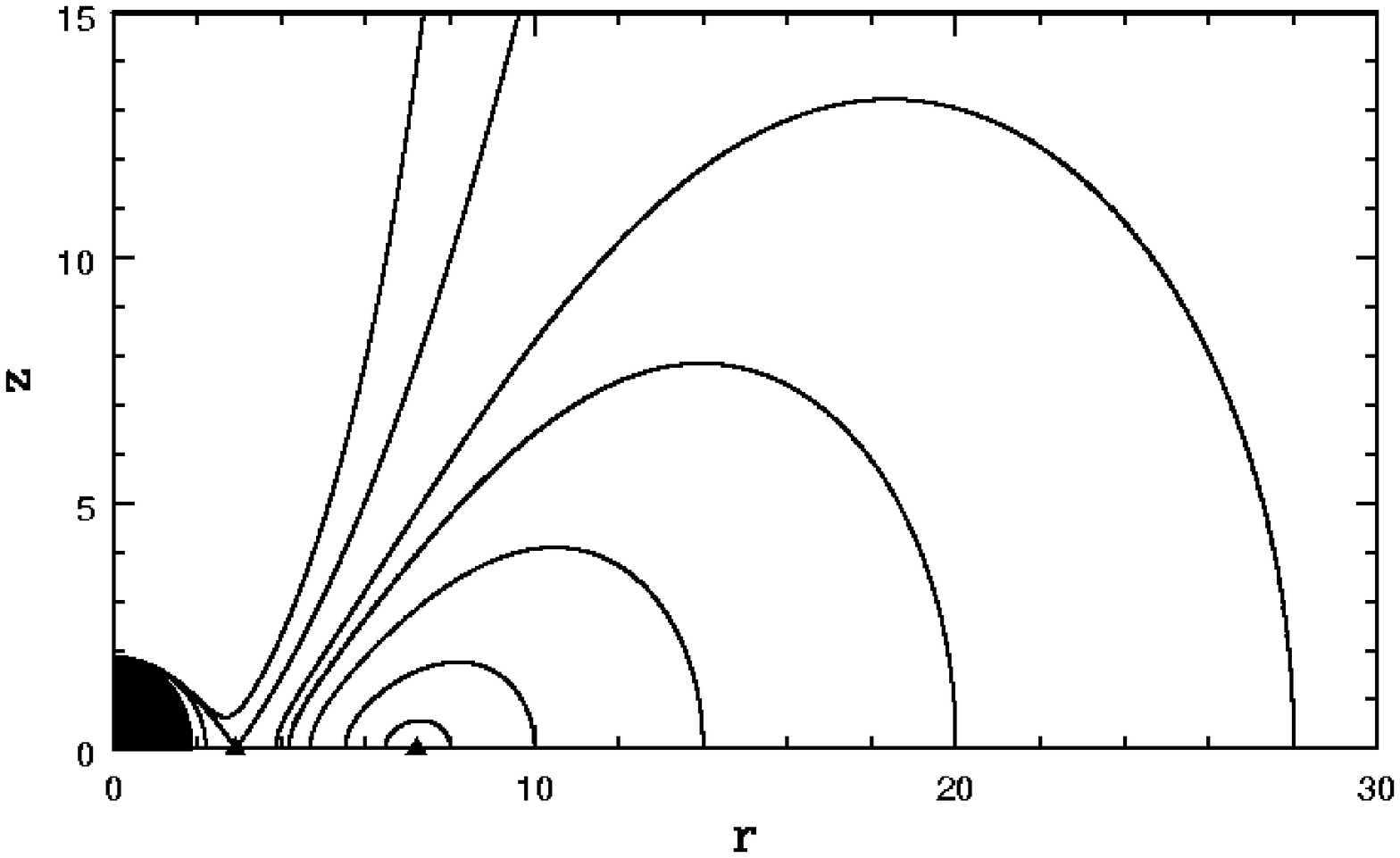}
   \includegraphics[width=4.3cm]{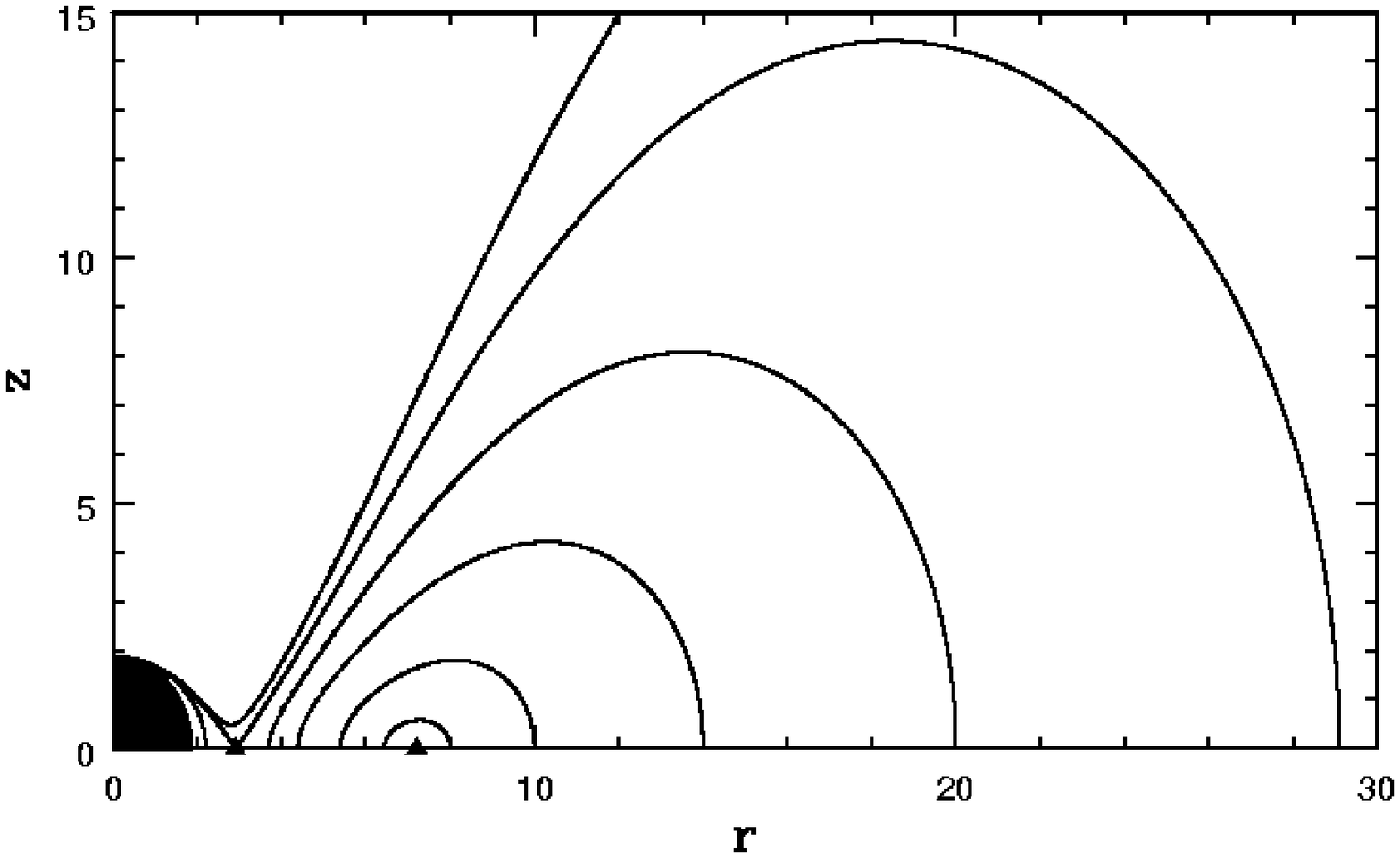}
   \includegraphics[width=4.3cm]{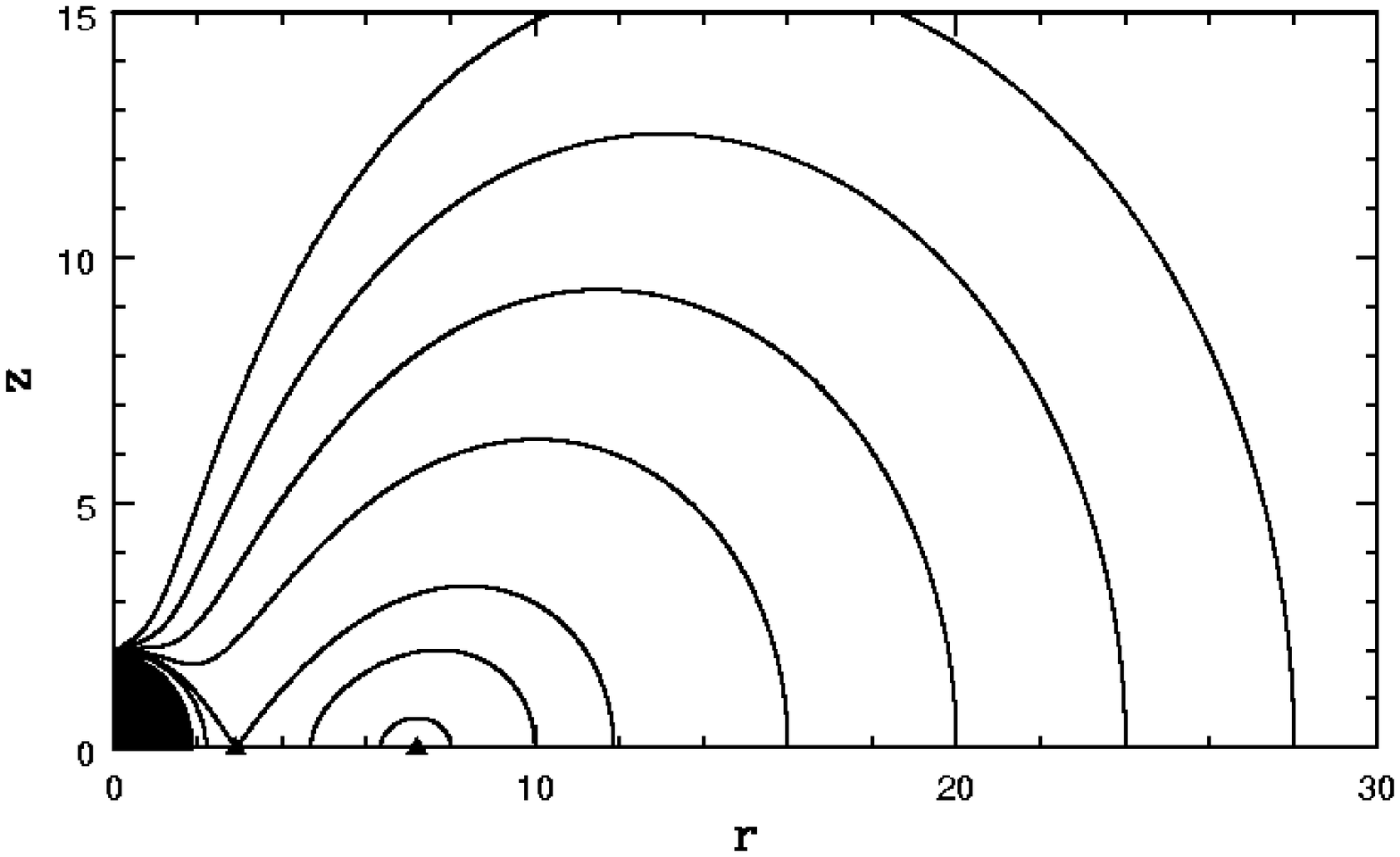}
   \includegraphics[width=4.3cm]{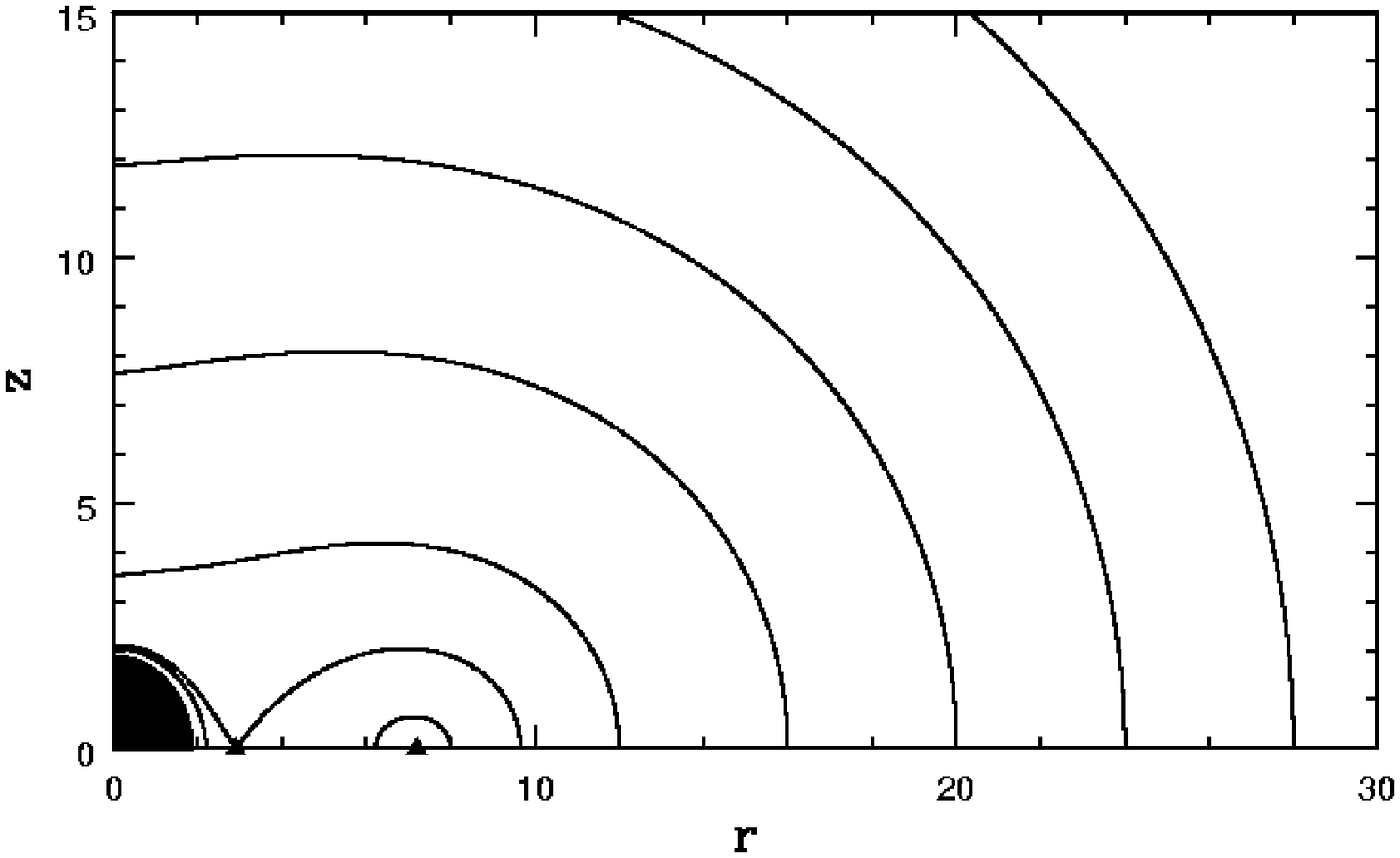}
   \includegraphics[width=4.3cm]{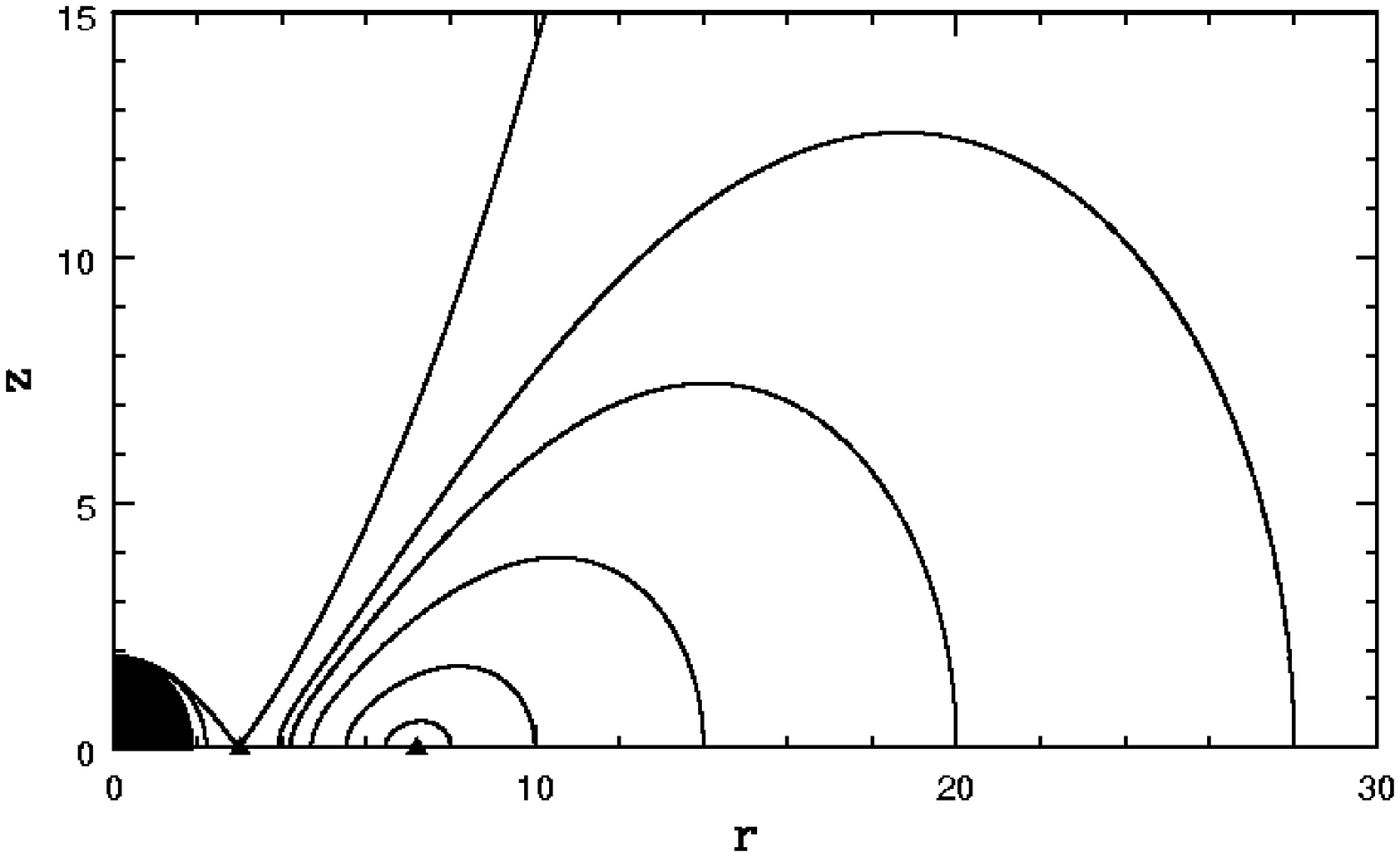}
   \includegraphics[width=4.3cm]{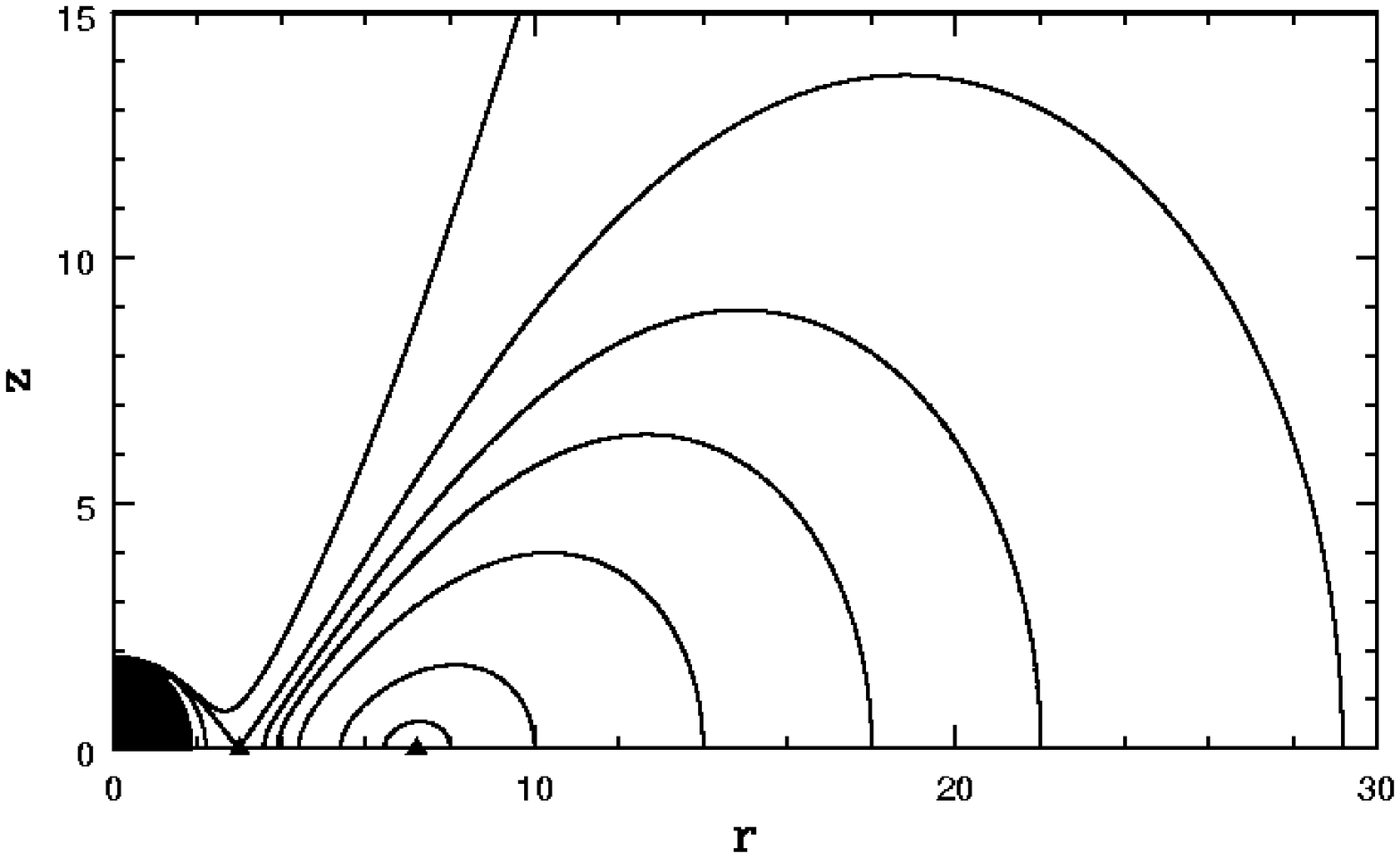}
   \includegraphics[width=4.3cm]{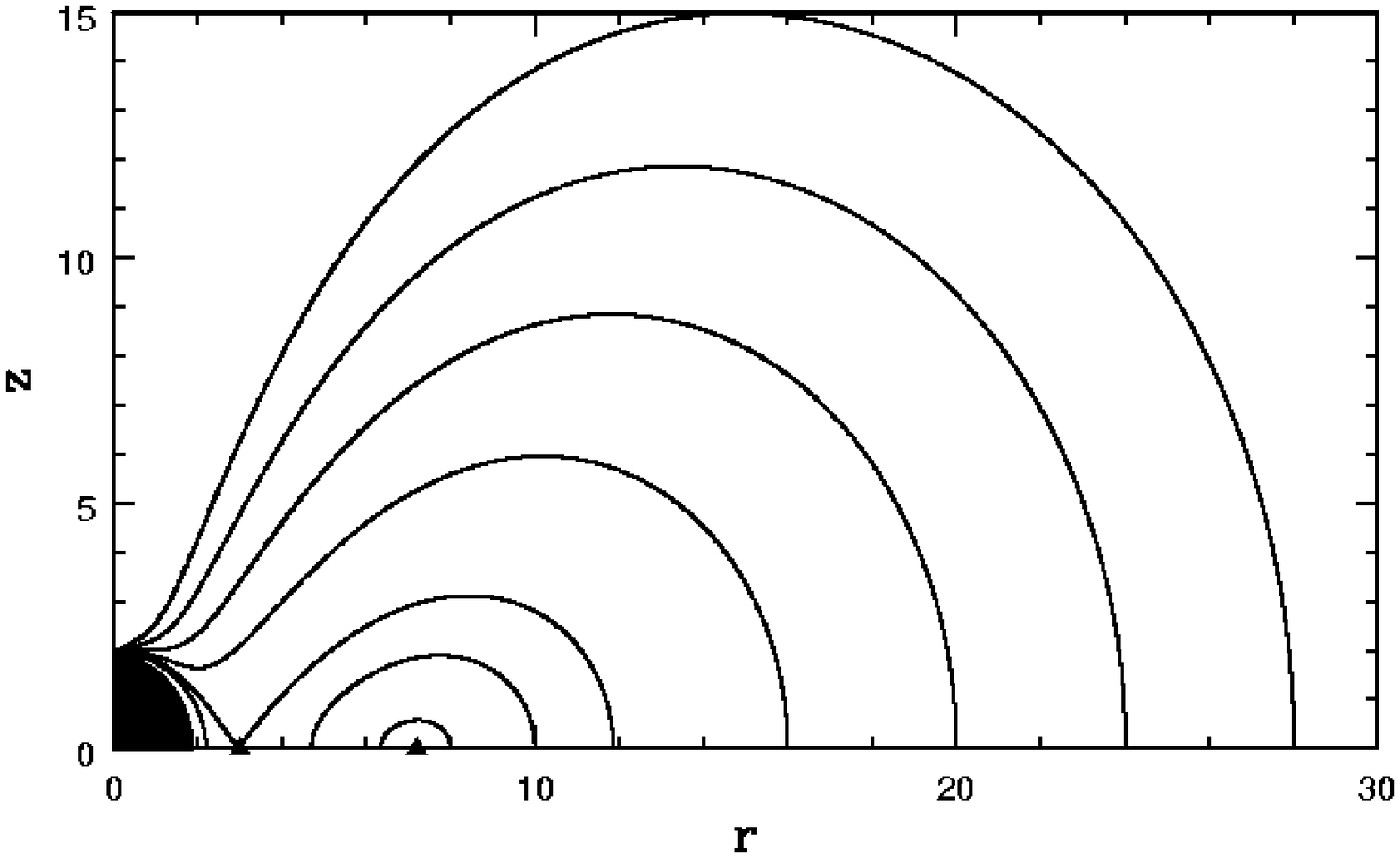}
   \includegraphics[width=4.3cm]{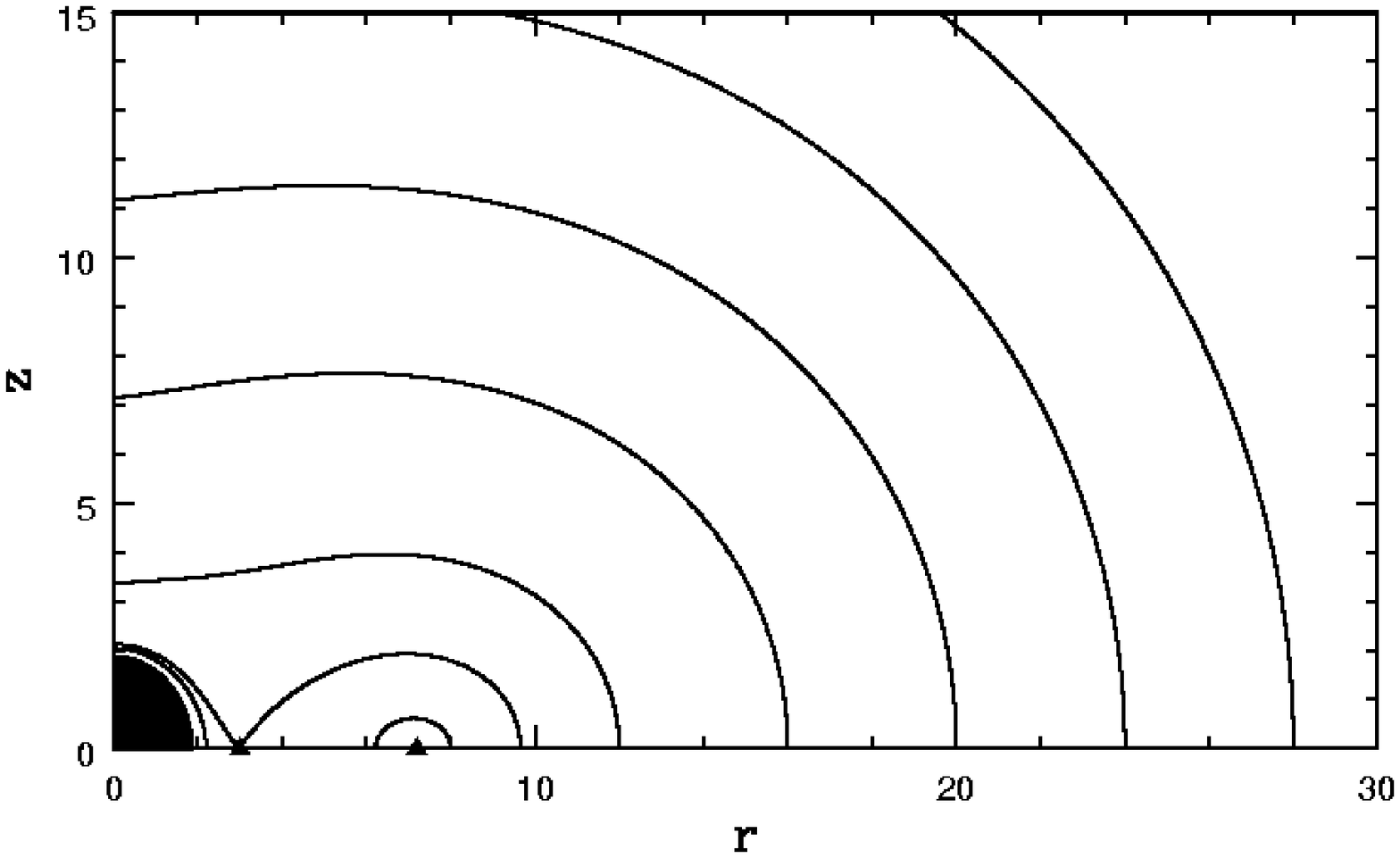}
   \includegraphics[width=4.3cm]{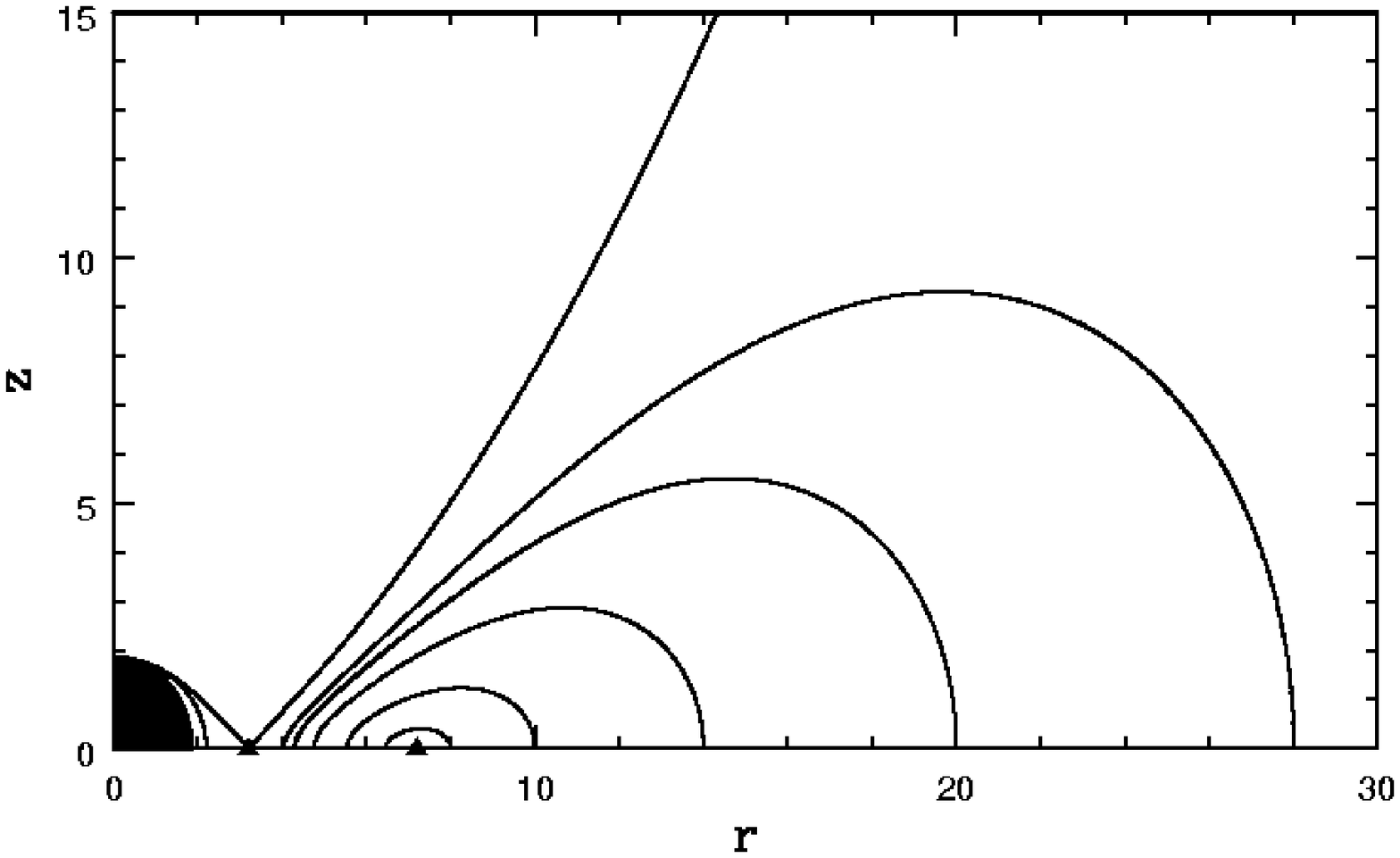}
   \includegraphics[width=4.3cm]{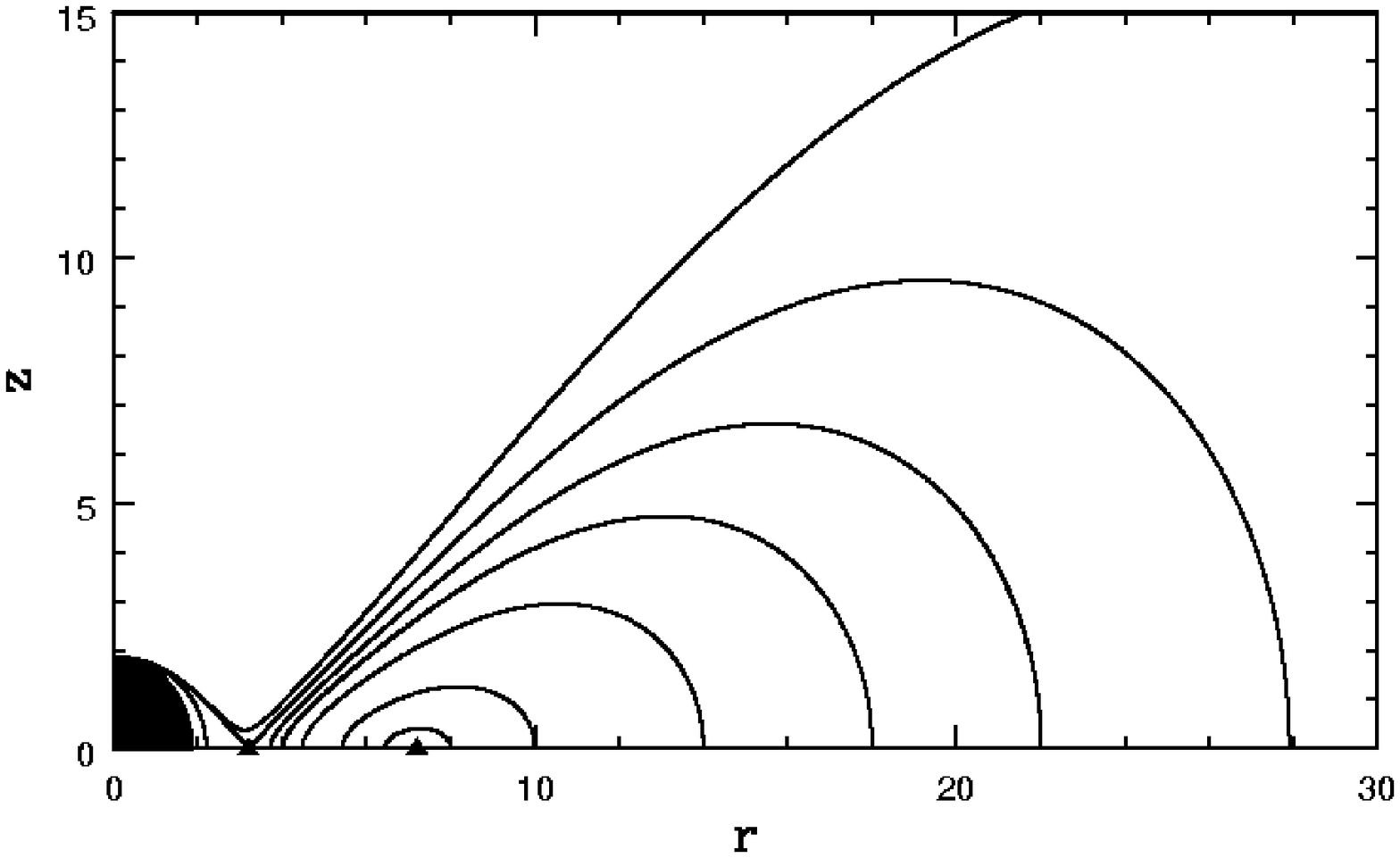}
   \includegraphics[width=4.3cm]{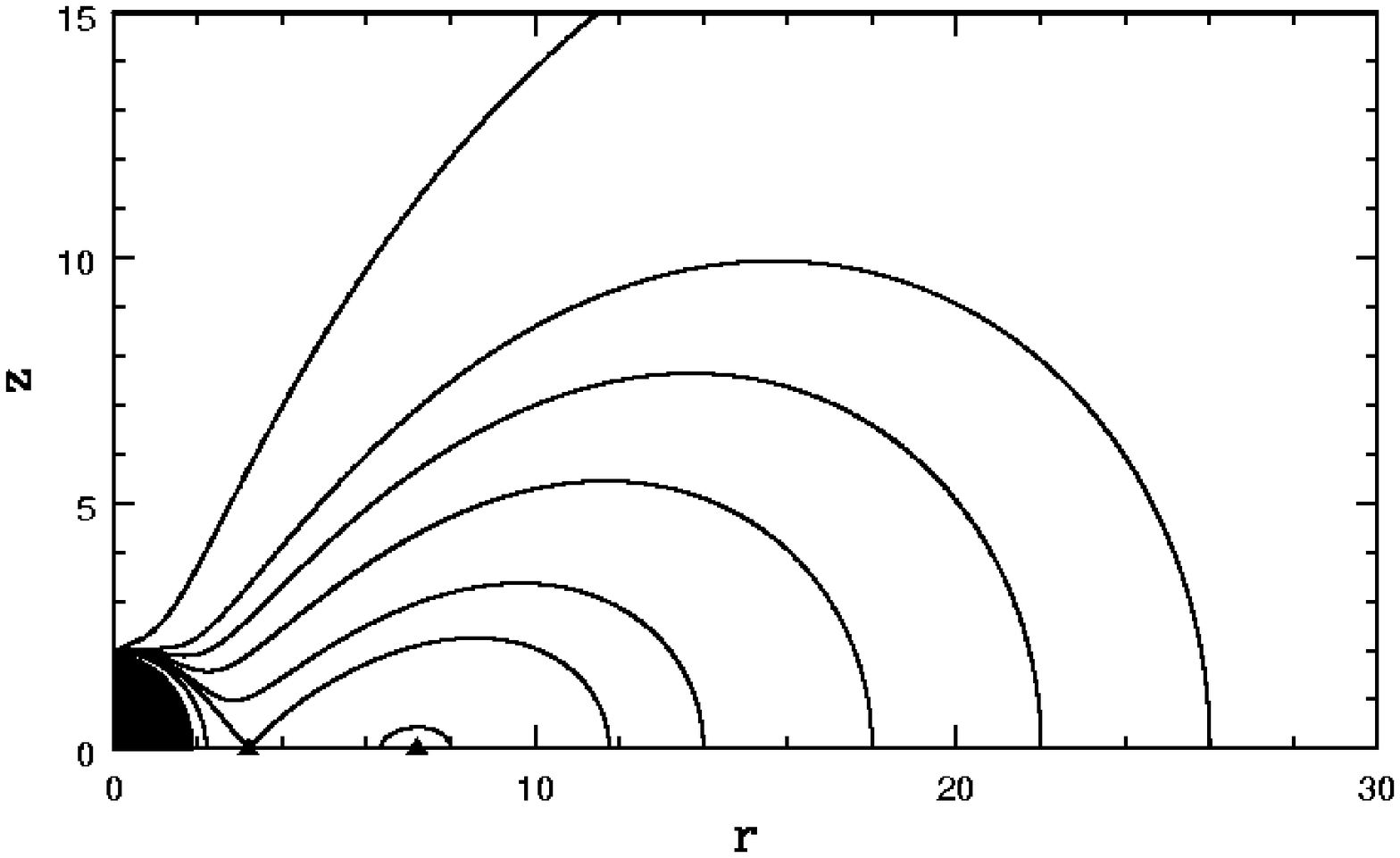}
   \includegraphics[width=4.3cm]{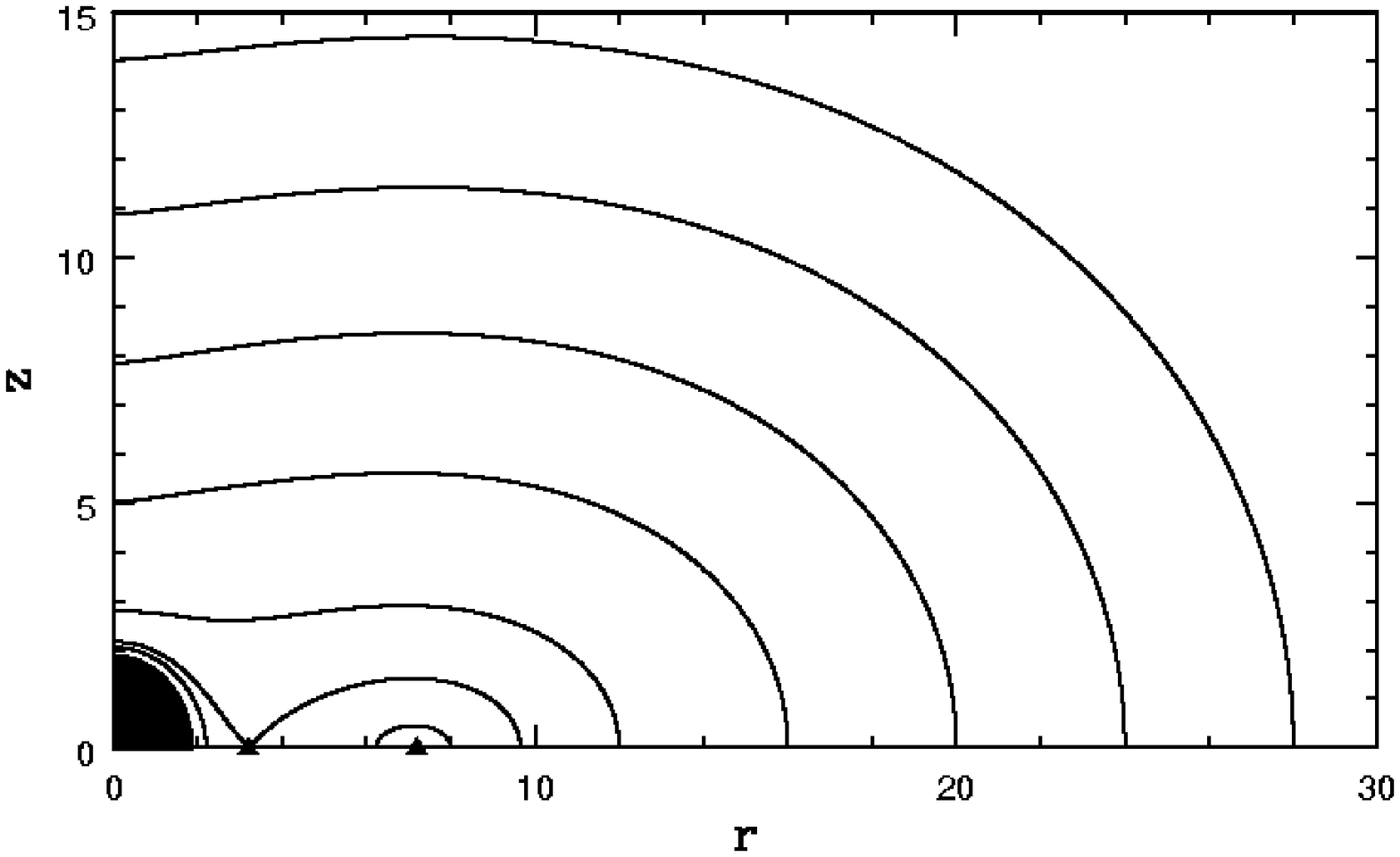}
   \includegraphics[width=4.3cm]{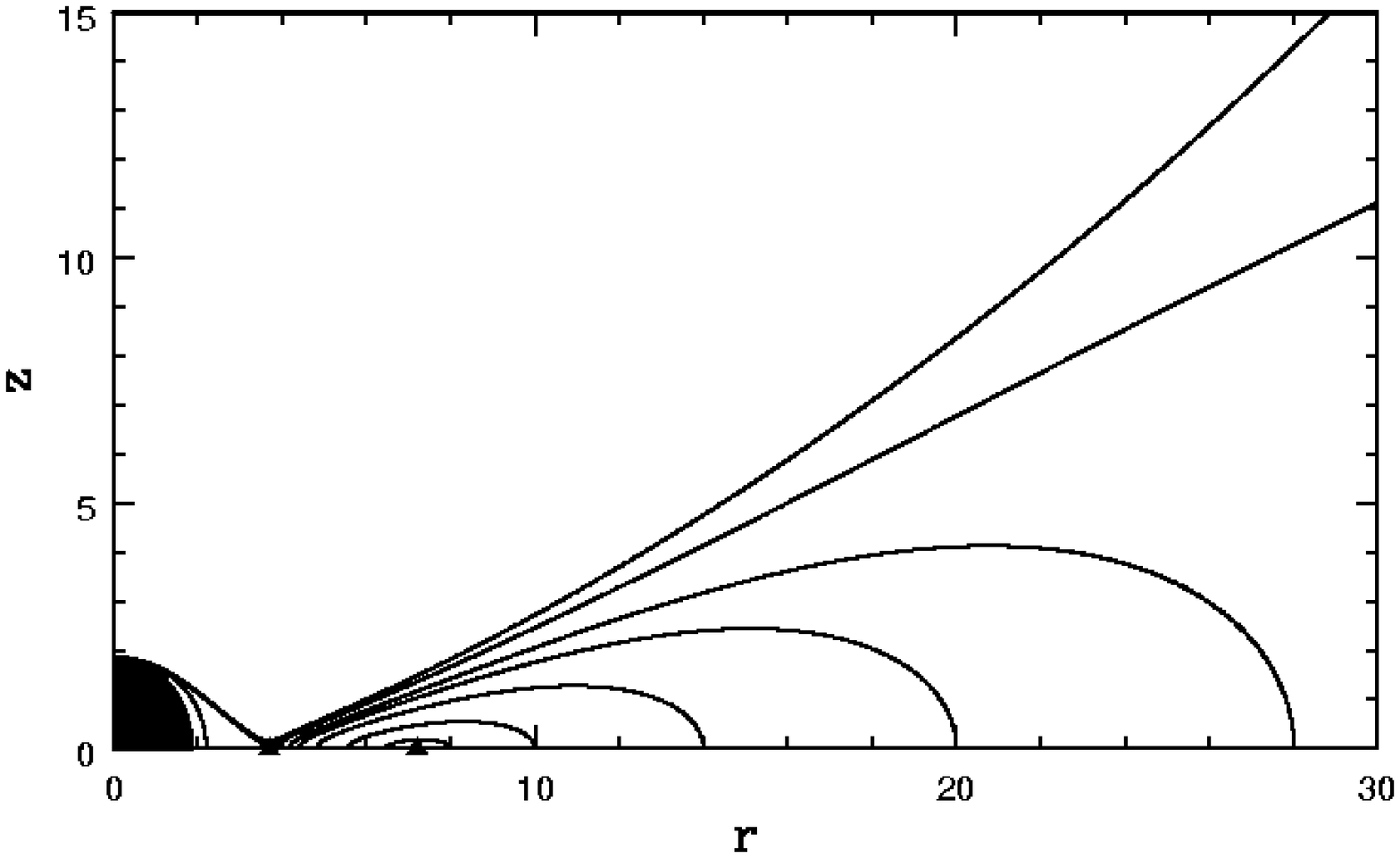}
   \includegraphics[width=4.3cm]{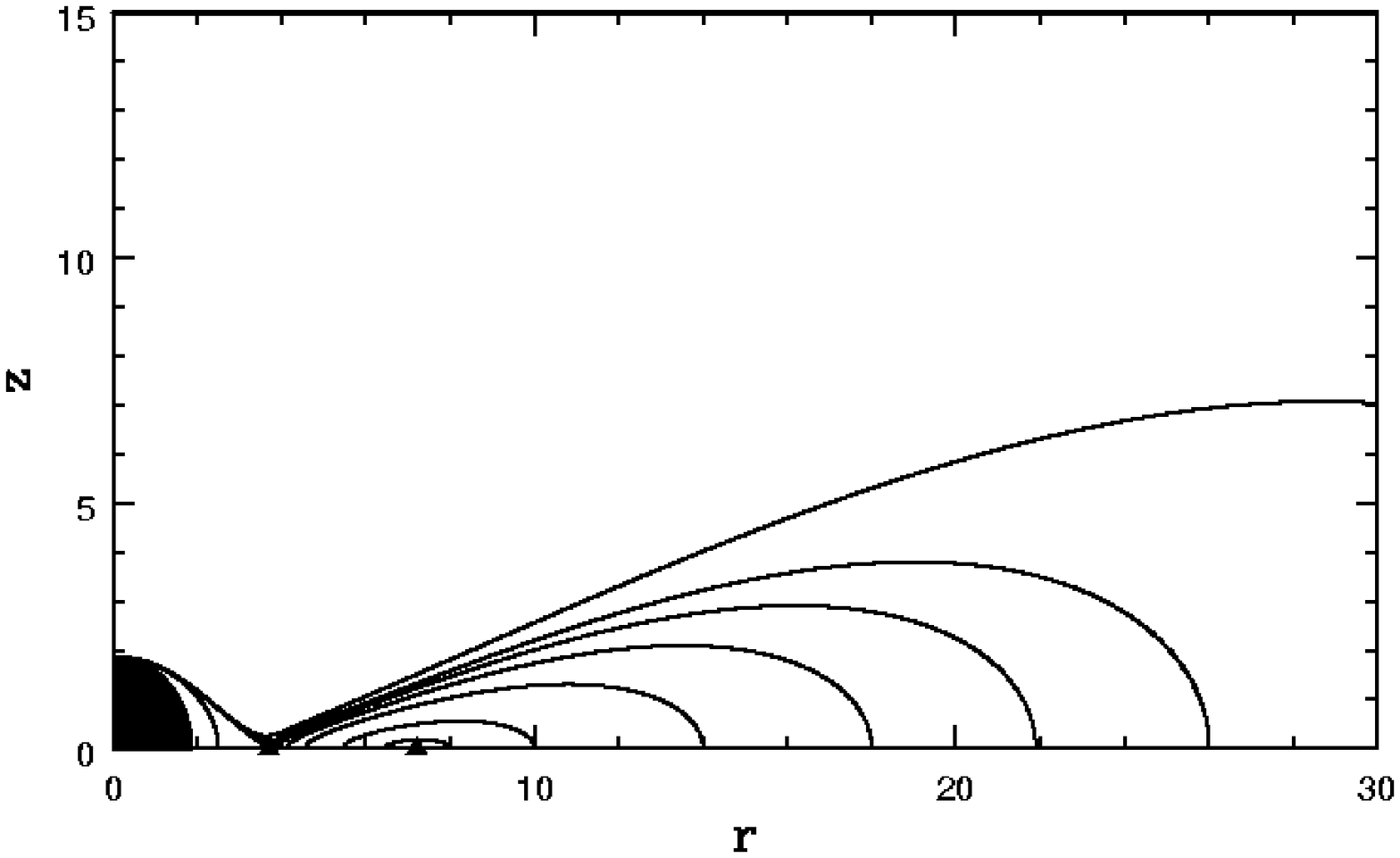}
   \includegraphics[width=4.3cm]{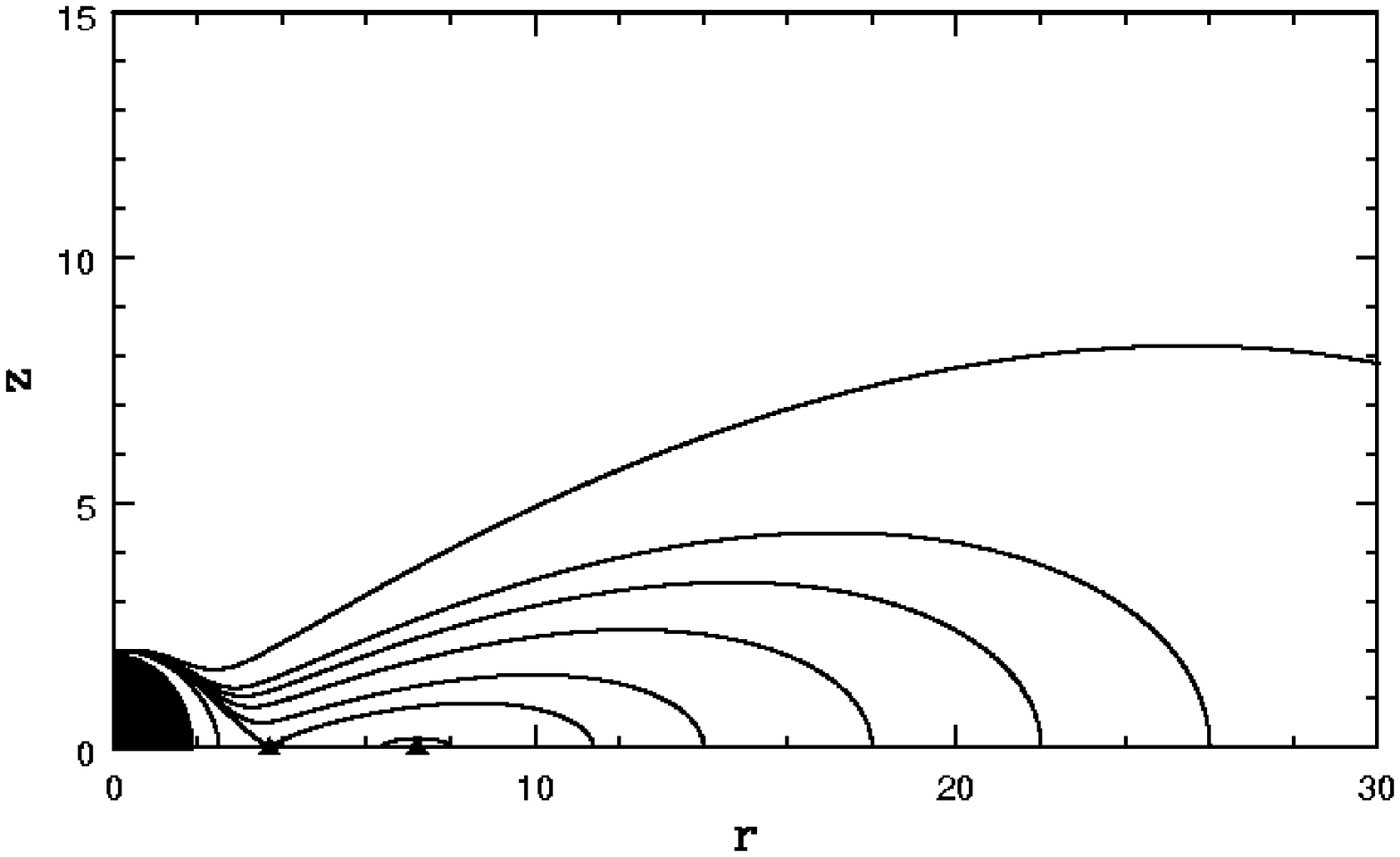}
   \includegraphics[width=4.3cm]{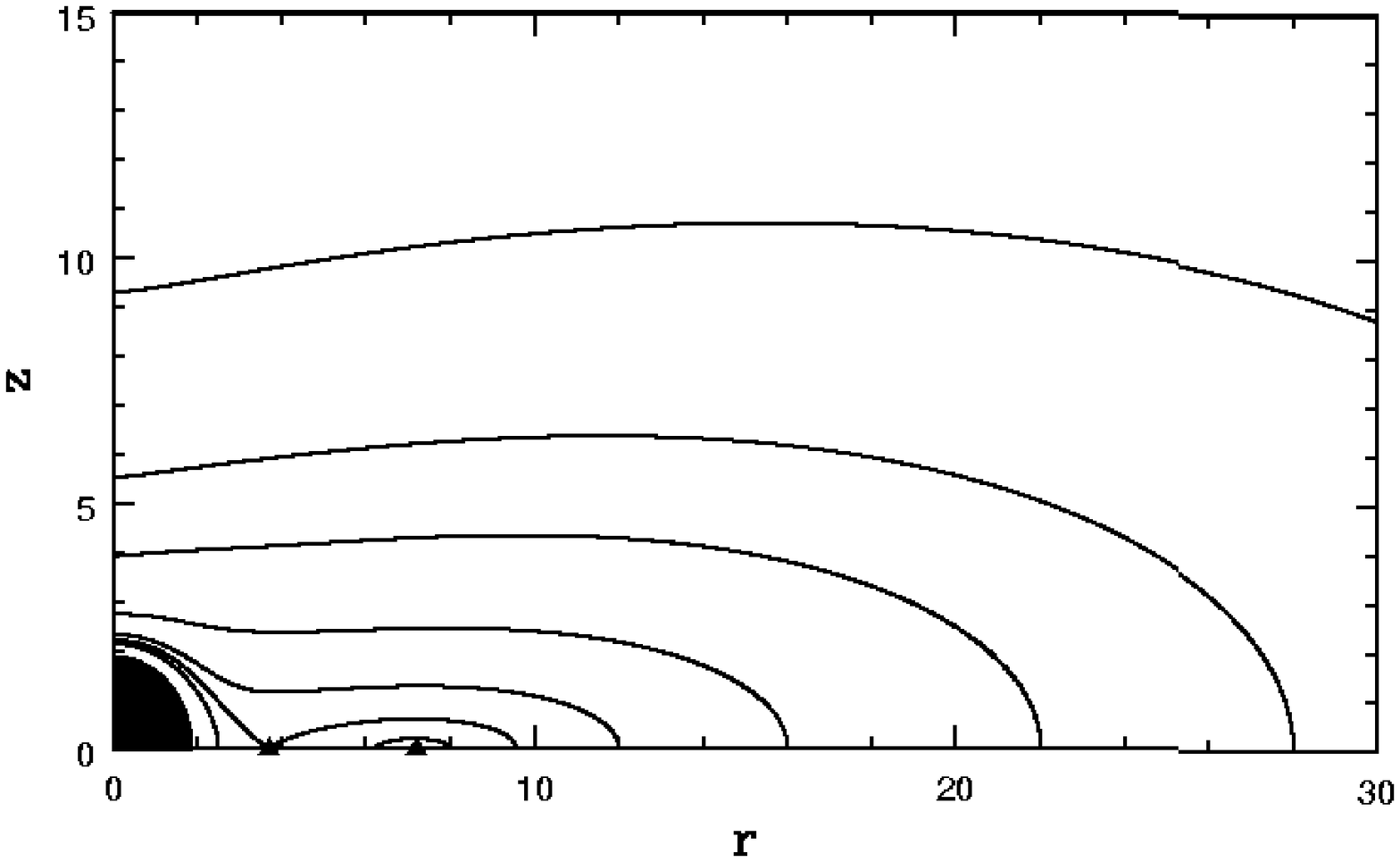}
   \includegraphics[width=4.3cm]{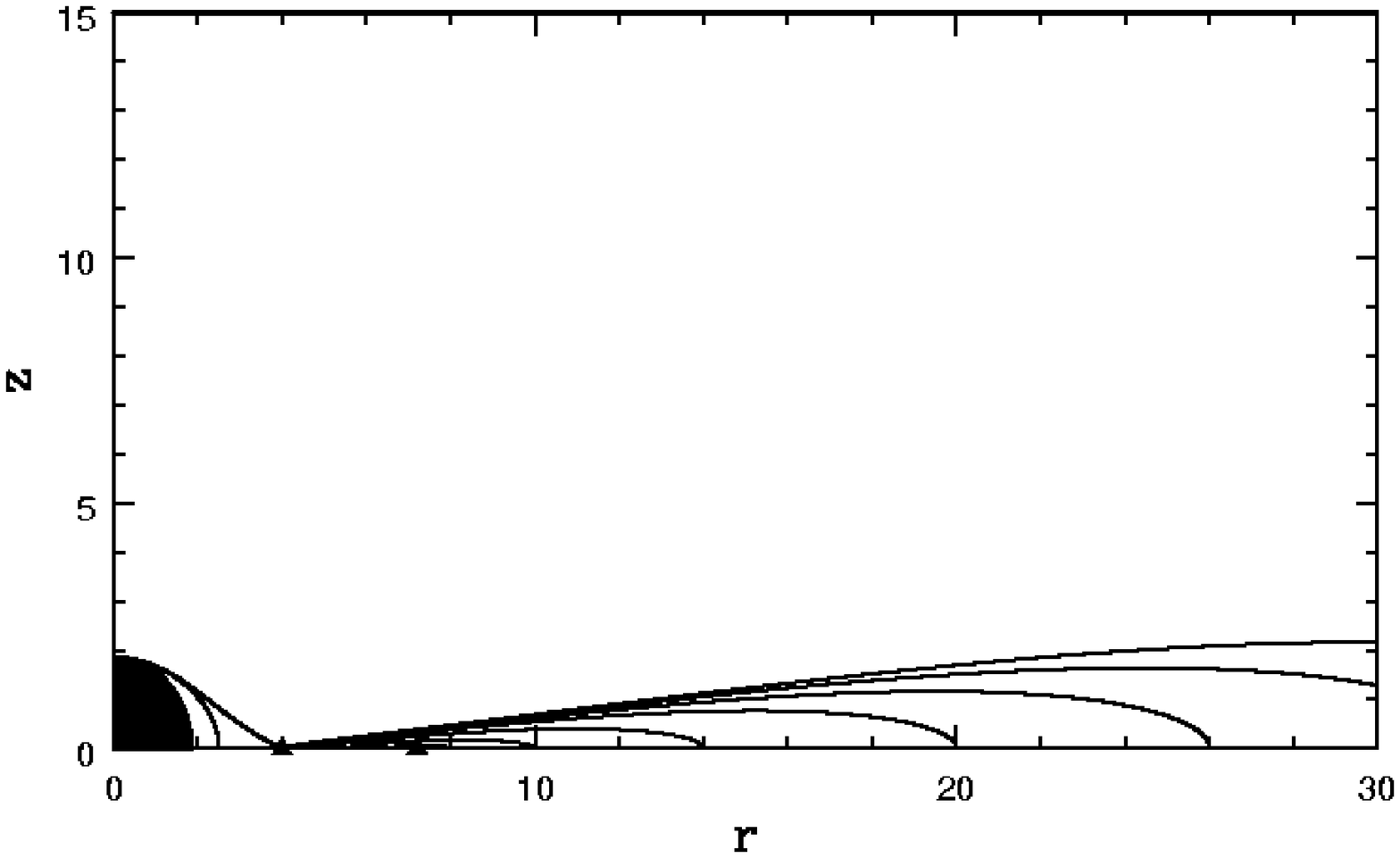}
   \includegraphics[width=4.3cm]{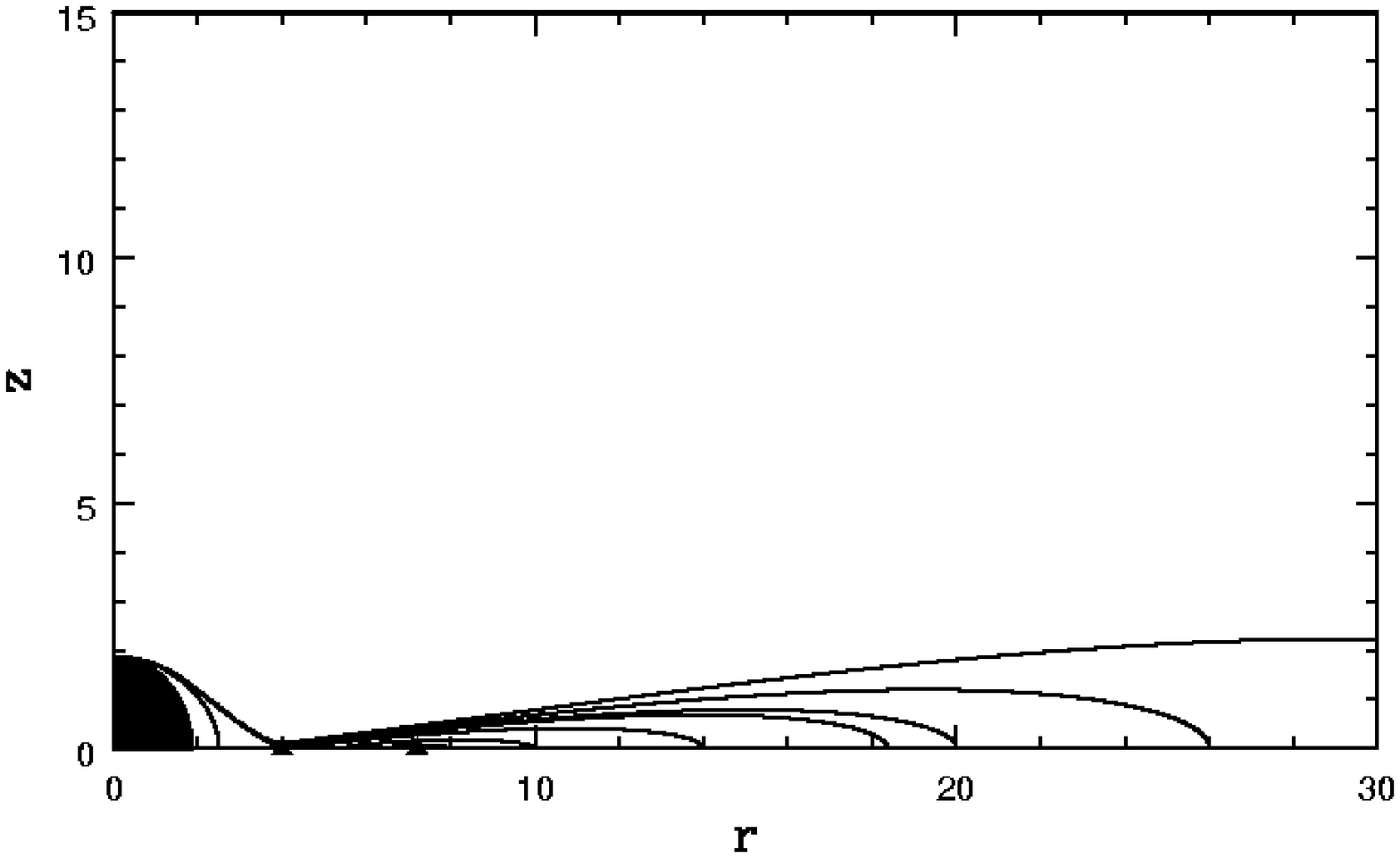}
   \includegraphics[width=4.3cm]{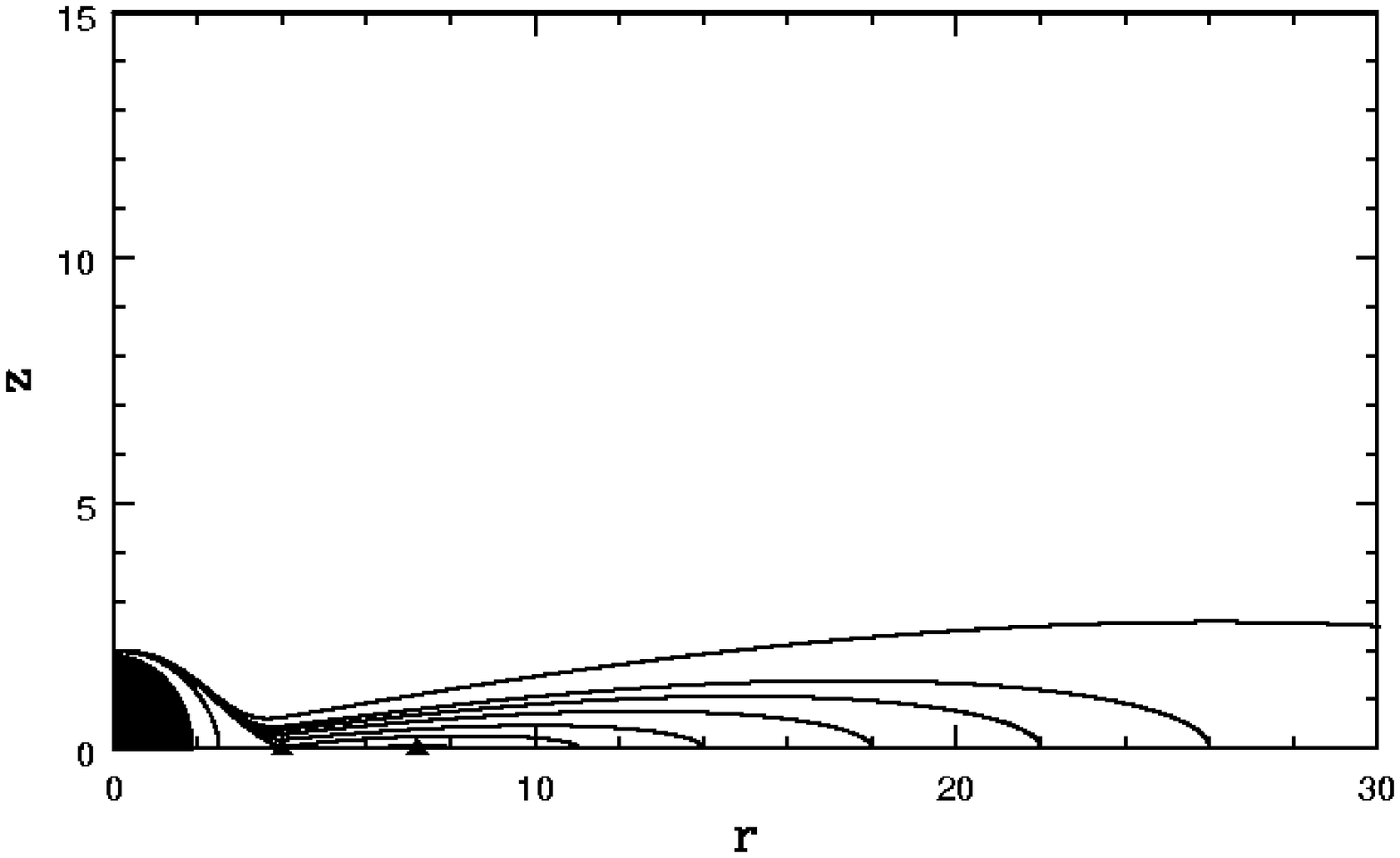}
   \includegraphics[width=4.3cm]{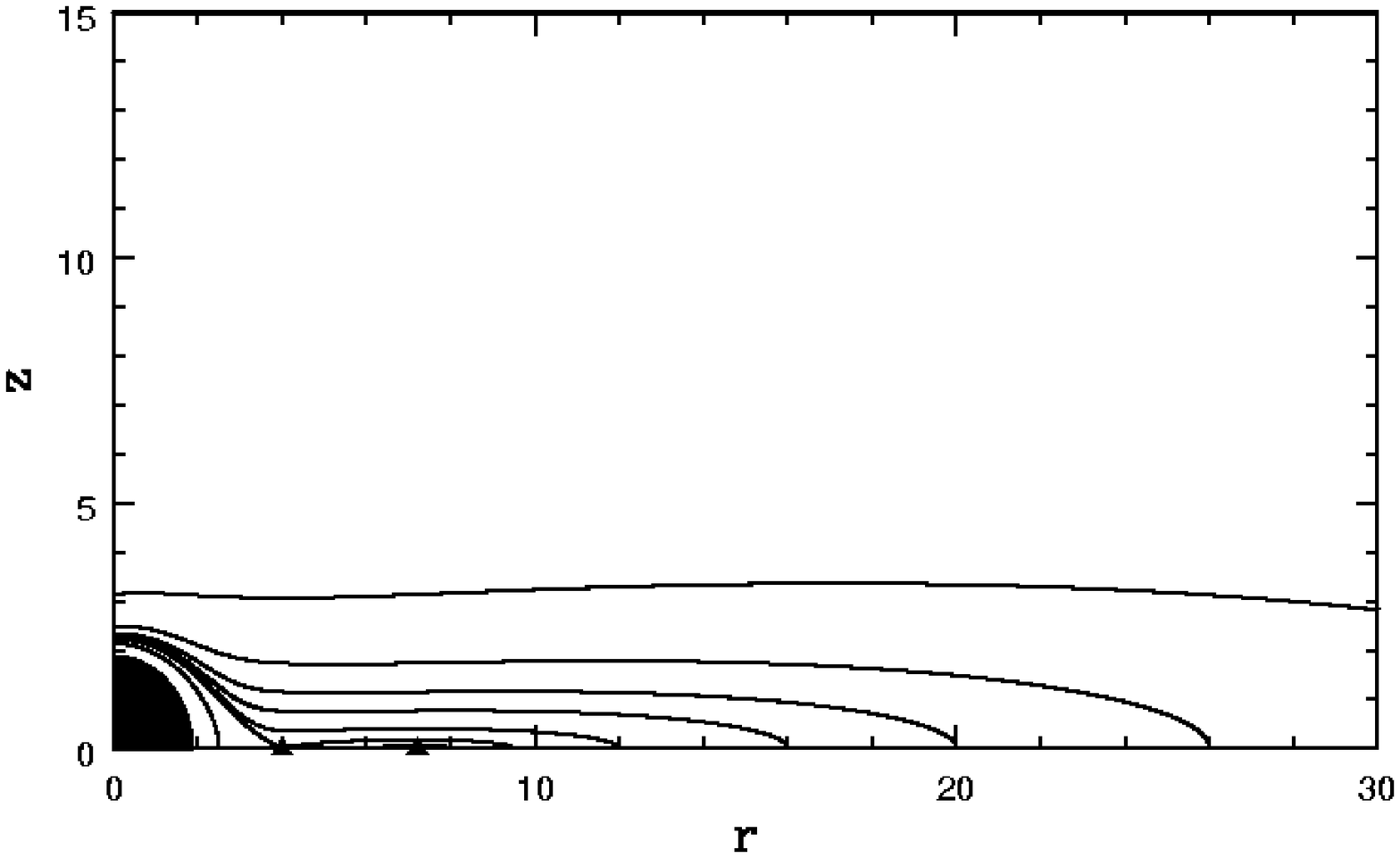}
      \caption
      {Equipressure surfaces for $a = 0.5$ and $\eta = \eta_{max} = 1.079$.
      Five rows correspond to $\beta = (0.0), (0.1), (0.5), (0.9), (0.99)$
      from the top to the bottom.
      Four columns correspond to $\gamma = (0.0), (0.1), (0.5), (0.9)$
      from the left to the right.
      The upper left corner shows a ``standard'' Polish doughnut. The
      lower right corner shows an almost Keplerian disk at the
      equatorial plane, surrendered by a very low angular momentum
      envelope.
      }
         \label{sequence-beta-gamma-05}
   \end{figure*}

Figures \ref{sequence-beta-gamma} and \ref{sequence-beta-gamma-05}
show sequences of models calculated with the new ansatz
(\ref{ansatz-general}) for black-hole spins $a=0$ and 0.5,
respectively. For these models we hold $\eta = \eta_{max}$ fixed,
while $\beta$ and $\gamma$ are varied over the limits of their
accessible ranges.

   \subsection{Equipressure surfaces on the axis of rotation}

Figures \ref{sequence-beta-gamma} and \ref{sequence-beta-gamma-05}
show an interesting change of
the behavior of equipressure surfaces close to the axis with
increasing $\gamma$. No equipressure surface can cross the symmetry
axis when the dependence of the angular momentum on $\theta$ is
weak. This is the case for the first three columns of
Figure \ref{sequence-beta-gamma} where $\gamma\leq 0.5$. On the
other hand, for the angular momentum distributions with higher
$\gamma$ the equipressure surfaces cross the axis perpendicularly.
This happens in plots of the last column of Figure
\ref{sequence-beta-gamma}. This behavior can be understood easily
from the limit of $r d\theta/dr$ as $\theta\rightarrow0$. In
Schwarzschild spacetime equations (\ref{master}) and
(\ref{differential}) give
\begin{equation}
  \label{limit}
  \lim_{\theta\rightarrow0^{+}}\frac{r d\theta}{dr} =
  -\frac{2\mathcal{L}^2_K(r)}{\mathcal{L}^2(r,\pi/2)}
  \lim_{\theta\rightarrow0^{+}}\left(\sin^{4\gamma-3}\theta\right).
\end{equation}
The limit on the right-hand side is either 0, 1 or $\infty$,
depending on the value of $\gamma$. When $\gamma<3/4$,
$rd\theta/dr =0$ and no equipressure surface goes across the axis.
On the other hand, when $\gamma>3/4$ equipressure surfaces cross
the axis perpendicularly. 
Of course, a stationary torus may exist only within an
equipotential surface located inside the Roche lobe,
i.e., the critical self-crossing equipotential within the cusp 
\citep[][]{abr-1985}.

   \subsection{Comparison with numerical simulations}

\begin{figure}
  \centering
  \vskip 0.2truecm
   \includegraphics[width=8.0cm]{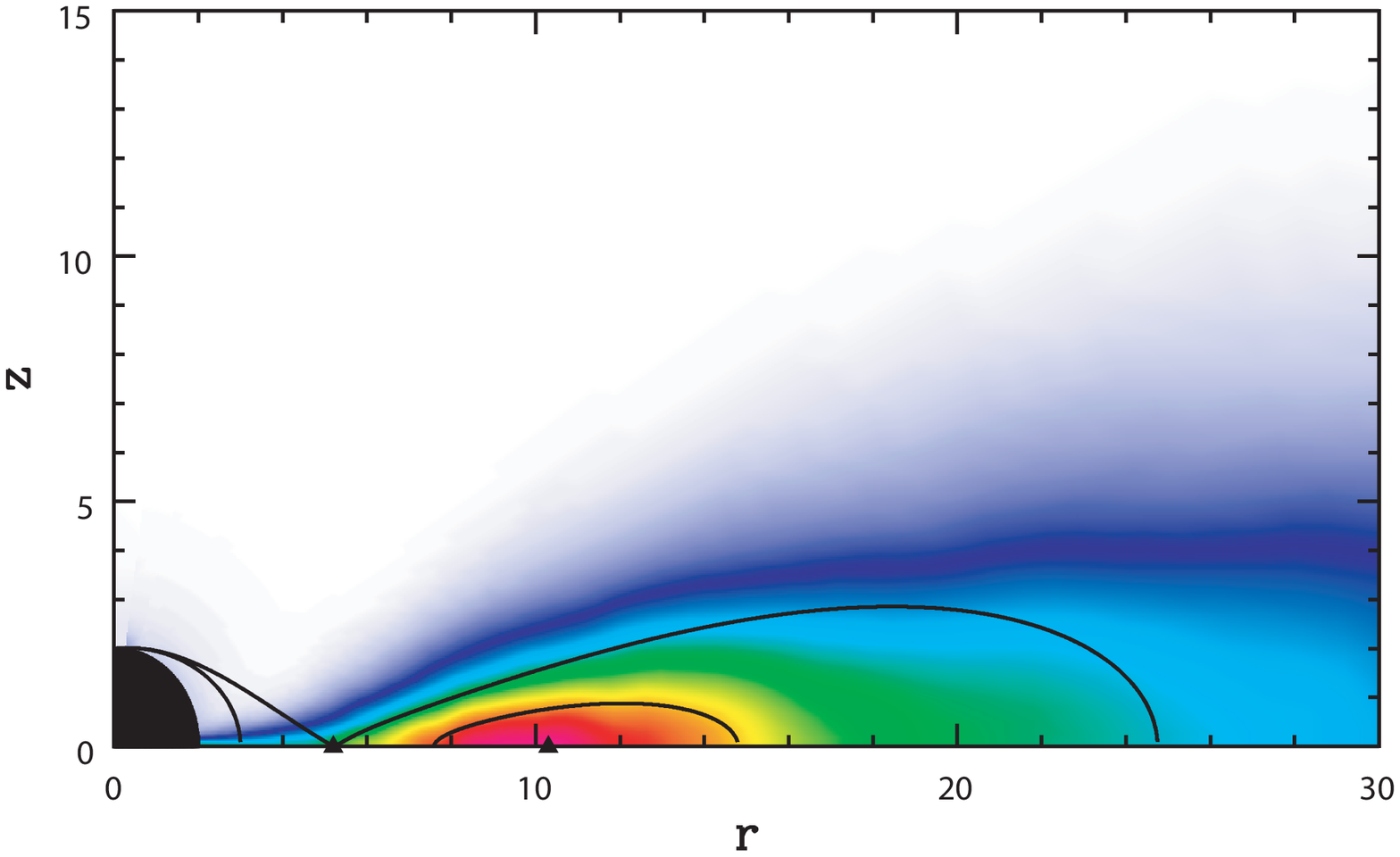}
   \vskip 0.8truecm
   \includegraphics[width=8.0cm]{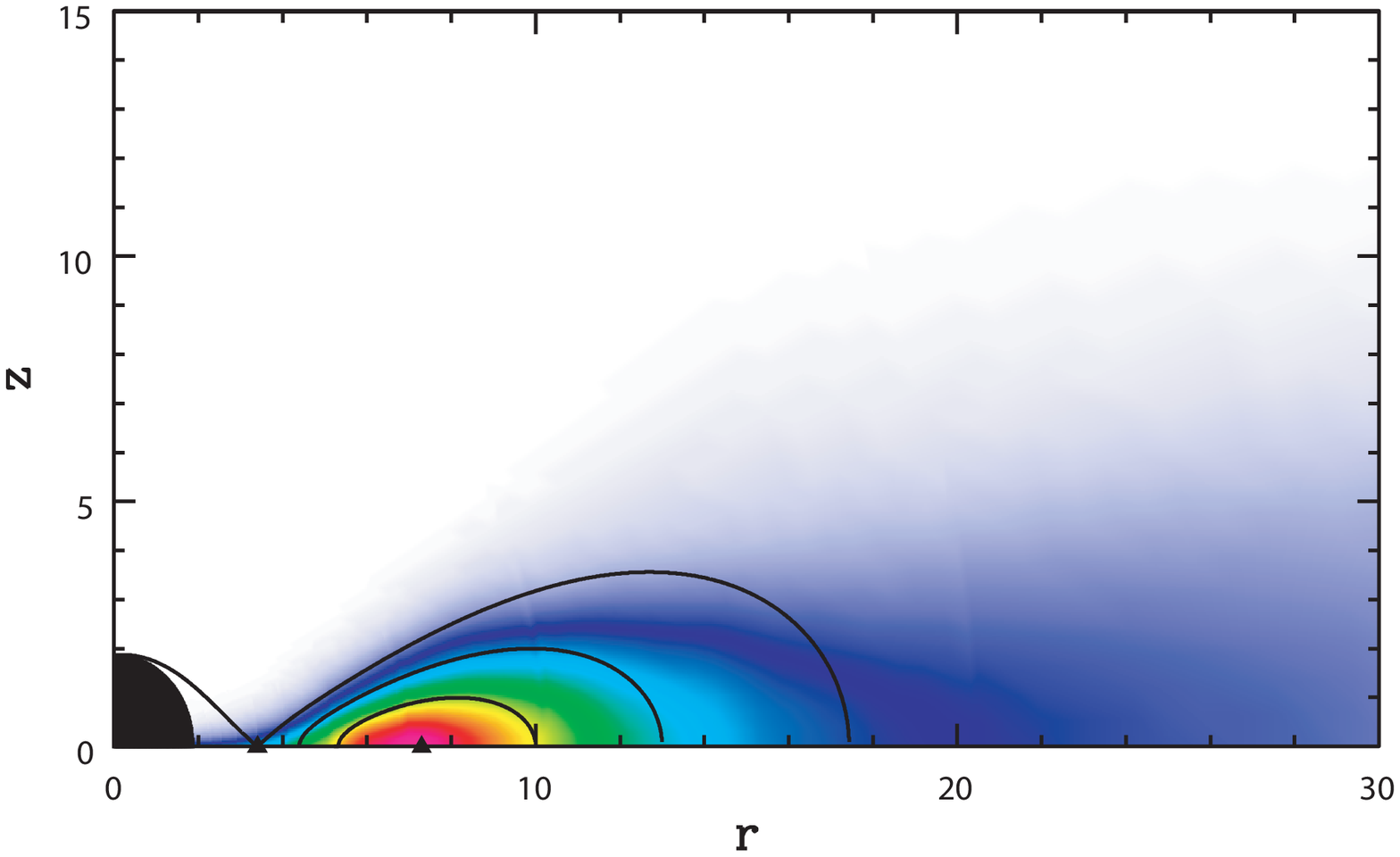}
      \caption
      {Comparison of pressure distributions between the analytic model ({\it dark lines})
      and numerical simulations ({\it colors}). The results of MHD
      simulations \citep[taken from][]{fra-2007,fra-2008} have been
      time-averaged over one orbital period at $r=25r_G$.
      {\it Upper panel:} Schwarzschild black hole ($a=0$); the analytic model
       parameters are $\eta=1.085$, $\beta=0.9$, and
       $\gamma=0.18$. {\it Lower panel:} Kerr black hole ($a=0.5$); the analytic model
       parameters are $\eta=1.079$, $\beta=0.7$, and
       $\gamma=0.2$.
      }
         \label{overlay}
   \end{figure}

Figure \ref{overlay} illustrates that the results of the analytic
models are well matched with results of modern 3-D MHD numerical
simulations \citep[here taken from][]{fra-2007,fra-2008}. For the
correct choice of parameters, the model can reproduce many of the
relevant features of the numerical results, including the
locations of the cusp and pressure maximum, as well as the
vertical thickness of the disk. At this stage, such qualitative
agreement is all that can be hoped for. One notable difference
between the analytic and numerical solutions is the behavior
inside the cusp. While the analytic equipressure surfaces formally
diverge toward the poles, the numerical solution maintains a
fairly constant vertical 
height, which is also evident in Figure \ref{analytic-numerical}.
This is because in the region inside the cusp, our assumption (\ref{circular-orbits}) about the form of the velocity is not valid --- velocity cannot be consistent with a pure rotation only, $u^i = (u^t, u^{\phi}, 0, 0)$. In this region the radial velocity $u^r$ must be non-zero and large. Thus, accuracy of our analytic models may only be trusted in the region outside the cusp, $r > r_{cusp}$.


\section{Discussion}
\label{discussion}


In this paper we assumed a form of the angular momentum
distribution (\ref{ansatz-general}) and from this calculated the
shapes and locations of the equipressure surfaces. This may be
used in calculating spectra (in the optically thick case) by the
same ``surface'' method as used in works by \citet{sik-1971} and
\citet{mad-1988}.

We plan to construct the complete physical model of the interior
in the second paper of this series. Here, we only outline the
method by considering a simplified toy model. Let us denote $\rho
= \epsilon + p$. We assume a toy (non-barytropic) equation of
state and an entropy distribution, by writing,
\begin{equation}
  p = e^{K({\cal S})}\rho, \quad
  K = K(r, \theta).
  \label{toy-state}
\end{equation}
Let us, in addition, define two functions connected to the entropy
distribution,
\begin{equation}
  \partial_\theta\,K =\kappa(r, \theta).
  \quad
  \frac{\partial_r K}{\partial_{\theta} K} =
  \lambda(r, \theta),
  \label{two-functions-entropy}
\end{equation}
From the obvious condition that the second derivative commutator
of pressure vanishes, $(\partial_r\partial_{\theta} -
\partial_{\theta}\partial_r)p = 0$, and equations (\ref{toy-state}),
(\ref{two-functions-entropy}) and (\ref{master}) one derives,
\begin{equation}
  \kappa = -\frac{\partial_r\,G_{\theta} -
  \partial_{\theta}\,G_r}{G_r - \lambda\,G_{\theta}},
  \label{commutator-condition}
\end{equation}
where $G_r$ and $G_{\theta}$ are defined as
\begin{equation}
  G_i(r,\theta) = \frac{\partial_i p}{\rho}
  \label{definition-G}
\end{equation}
and can be calculated from the angular momentum distribution using
equation (\ref{Euler}). From (\ref{commutator-condition}) it is
obvious that one cannot independently assume the functions
$\kappa(r, \theta)$ and $\lambda(r, \theta)$\footnote{A somewhat
similar situation in the case of rotating stars is known as the
von Zeipel paradox \citep{tas-1978}: {\it Pseudo-barytropic models
in a state of permanent rotation cannot be used to describe
rotating stars in strict radiative equilibrium.}}. Assuming
$\lambda(r, \theta)$ is equivalent with assuming the shapes of
isentropic surfaces. Indeed, from (\ref{two-functions-entropy})
one concludes that the function $\theta = \theta_{\cal S}(r)$ that
describes an isentropic surface is given by the equation,
\begin{equation}
  \left[\frac{d\theta}{dr}\right]_{\cal S} =
  -\lambda(r, \theta).
 \label{isentropic}
\end{equation}
that may be directly integrated. Then the condition
(\ref{commutator-condition}) gives the physical spacing
(``labels'') to the isentropic surfaces, and through the equation
of state (\ref{toy-state}) also to equipressure surfaces and
isopicnic  ($\rho ={\rm const}$) surfaces.

Note, that a possible choice $\lambda = G_r/G_{\theta}$
corresponds, obviously, to the ``von Zeipel'' case in which
equipressure and isentropic surfaces coincide. In this case the
denominator in (\ref{commutator-condition}) vanishes, implying a
singularity unless the numerator also vanishes. The condition for
the numerator to vanish is, however, equivalent to the von Zeipel
condition.


\section{Conclusions}


The new ansatz (\ref{ansatz-general}) captures two essential
features of the angular momentum distribution in black hole
accretion disks:
   \begin{enumerate}
      \item On the equatorial plane and far from the black hole,
      the angular momentum in the disk differs only little from the
      Keplerian one being slightly sub-Keplerian, but closer in
      it becomes (slightly) super-Keplerian and still closer, in the
      plunging region, sub-Keplerian again and nearly constant.
      \item Angular momentum may significantly decrease off the
      equatorial plane, and become very low (even close to zero,
      in a non-rotating ``corona'').
   \end{enumerate}
Models of tori described here may be useful not only for accretion disks but also for tori that form in the latest stages of neutron star binary mergers. This is relevant for gamma ray bursts \citep{wit-1994} and gravitational waves \citep{bai-2008}.

\begin{acknowledgements}
We thank Daniel Proga and Luciano Rezzolla for helpful comments and suggestions.
Travel expenses connected to this work were 
supported by the China Scholarship Council (Q.L.), the Polish 
Ministry of Science grant N203 0093/1466 (M.A.A.), and the Swedish 
Research Council grant VR Dnr 621-2006-3288 (P.C.F.).
\end{acknowledgements}



\end{document}